\documentclass[twocolumn,showpacs,showkeys,preprintnumbers,amsmath,amssymb,superscriptaddress]{revtex4-1}
\usepackage{graphicx}
\usepackage{natbib}

\begin{document}
\typeout{Filename: reftest4-1.tex for revtex 4.1i 2009/10/19 (AO)}

\title{Characterization of the scission point from fission-fragment velocities}
\author{M.~Caama\~{n}o}
\email[Email address: ]{manuel.fresco@usc.es}
\affiliation{GANIL, CEA/DSM-CNRS/IN2P3, BP 55027, F-14076 Caen Cedex 5, France}
\affiliation{Universidade de Santiago de Compostela, E-15706 Santiago de Compostela, Spain}
\author{F.~Farget}
\email[Email address: ]{fanny.farget@ganil.fr}
\affiliation{GANIL, CEA/DSM-CNRS/IN2P3, BP 55027, F-14076 Caen Cedex 5, France}
\affiliation{Universidade de Santiago de Compostela, E-15706 Santiago de Compostela, Spain}
\author{O.~Delaune}
\altaffiliation[Present address:]{CEA DAM DIF, F-91297 Arpajon, France}
\affiliation{GANIL, CEA/DSM-CNRS/IN2P3, BP 55027, F-14076 Caen Cedex 5, France}
\author{K.-H.~Schmidt}
\affiliation{GANIL, CEA/DSM-CNRS/IN2P3, BP 55027, F-14076 Caen Cedex 5, France}
\author{C.~Schmitt}
\affiliation{GANIL, CEA/DSM-CNRS/IN2P3, BP 55027, F-14076 Caen Cedex 5, France}
\author{L.~Audouin}
\affiliation{IPN Orsay, IN2P3/CNRS-UPS, F-91406 Orsay Cedex, France}
\author{C.-O.~Bacri}
\affiliation{IPN Orsay, IN2P3/CNRS-UPS, F-91406 Orsay Cedex, France}
\author{J.~Benlliure}
\affiliation{Universidade de Santiago de Compostela, E-15706 Santiago de Compostela, Spain}
\author{E.~Casarejos}
\affiliation{Universidade de Vigo, E-36310, Spain}
\author{X.~Derkx}
\altaffiliation[Present address:]{Universit\"at Mainz, D-55128 Mainz, Germany}
\affiliation{GANIL, CEA/DSM-CNRS/IN2P3, BP 55027, F-14076 Caen Cedex 5, France}
\author{B.~Fern\'andez-Dom\'inguez}
\affiliation{University of Liverpool, Liverpool L69 7ZE, United Kingdom}
\author{L.~Gaudefroy}
\affiliation{CEA DAM DIF, F-91297 Arpajon, France}
\author{C.~Golabek}
\altaffiliation[Present address:]{CEA, Centre de Saclay, IRFU/SPhN, F-91191 Gif-sur-Yvette, France}
\affiliation{GANIL, CEA/DSM-CNRS/IN2P3, BP 55027, F-14076 Caen Cedex 5, France}
\author{B.~Jurado}
\affiliation{CENBG, IN2P3/CNRS-UB1, F-33175 Gradignan Cedex, France}
\author{A.~Lemasson}
\affiliation{GANIL, CEA/DSM-CNRS/IN2P3, BP 55027, F-14076 Caen Cedex 5, France}
\author{D.~Ramos}
\affiliation{Universidade de Santiago de Compostela, E-15706 Santiago de Compostela, Spain}
\author{C.~Rodr\'iguez-Tajes}
\affiliation{Universidade de Santiago de Compostela, E-15706 Santiago de Compostela, Spain}
\affiliation{GANIL, CEA/DSM-CNRS/IN2P3, BP 55027, F-14076 Caen Cedex 5, France}
\author{T.~Roger}
\affiliation{GANIL, CEA/DSM-CNRS/IN2P3, BP 55027, F-14076 Caen Cedex 5, France}
\author{A.~Shrivastava}
\affiliation{GANIL, CEA/DSM-CNRS/IN2P3, BP 55027, F-14076 Caen Cedex 5, France}
\affiliation{Nuclear Physics Division, Bhabha Atomic Research Centre, Mumbai 400085, India}
\date{\today}

\begin{abstract}
The isotopic-yield distributions and kinematic properties of fragments produced in transfer-induced fission of $^{240}$Pu and fusion-induced fission of $^{250}$Cf, with 9~MeV and 45~MeV of excitation energy respectively, were measured in inverse kinematics with the spectrometer VAMOS. The kinematic properties of identified fission fragments allow to derive properties of the scission configuration such as the distance between fragments, the total kinetic energy, the neutron multiplicity, the total excitation energy, and, for the first time, the proton- and neutron-number sharing during the emergence of the fragments. These properties of the scission point are studied as functions of the fragment atomic number. The correlation between these observables, gathered in one single experiment and for two different fissioning systems at different excitation energies, give valuable information for the understanding and modeling of the fission process.
\end{abstract}



\maketitle

\section{Introduction}

Nuclear fission is a complex process where a number of different properties of the nucleus influence the characteristics of the resulting splitting. In general, experimental observations are limited to few observables, making it difficult to isolate the impact of specific nuclear properties on the whole process. As an illustration, the fission-fragment mass yields have been measured intensively during the last decades, being in fact the result of the proton and neutron sharing between the two fragments. Isotopic yields have been accessible only for the light fragments~\cite{lang80,boc89} whereas techniques of delayed gamma-spectroscopy led to measures limited in range and in precision~\cite{bail11}. Few years back, the measure of the complete distribution of the fragment atomic number was feasible when based on techniques involving inverse kinematics~\cite{sch00}, leading to new observations that were not expected from the study of mass distributions~\cite{sch01, boeck08, stein98}. Recently, the use of inverse kinematics using transfer- and fusion-induced fission~\cite{caaPRC2013} has given access to the isotopic distributions of fragments issued from well-defined fissioning systems. In this scenario, the neutron excess (defined as the neutron-to-proton ratio) of each fragment over the whole distribution is a new observable that allows the investigation of the neutron and proton sharing between fragments during their emergence, the effect of shell structure at scission deformation, and the way both fragments share the total excitation energy. In order to reproduce the data, fission models need to make assumptions on the scission configuration (deformation and nucleon sharing) and the partition of excitation energy between both fragments. However, scission-point properties are very difficult to derive experimentally because the fragment detection occurs after neutron evaporation, hiding the exact neutron numbers of the fragments. In general, the elongation at scission of the fissioning system is revealed by the measure of the total kinetic energy~\cite{boeck08} while the deformation of fragments and the sharing of the excitation energy is derived from neutron multiplicities measured either directly~\cite{hinde92, nishio} or indirectly from the velocity and energy of both fragments measured in coincidence~\cite{mull84}.

\begin{figure*}[!t]
\begin{center}
\includegraphics[width=\columnwidth]{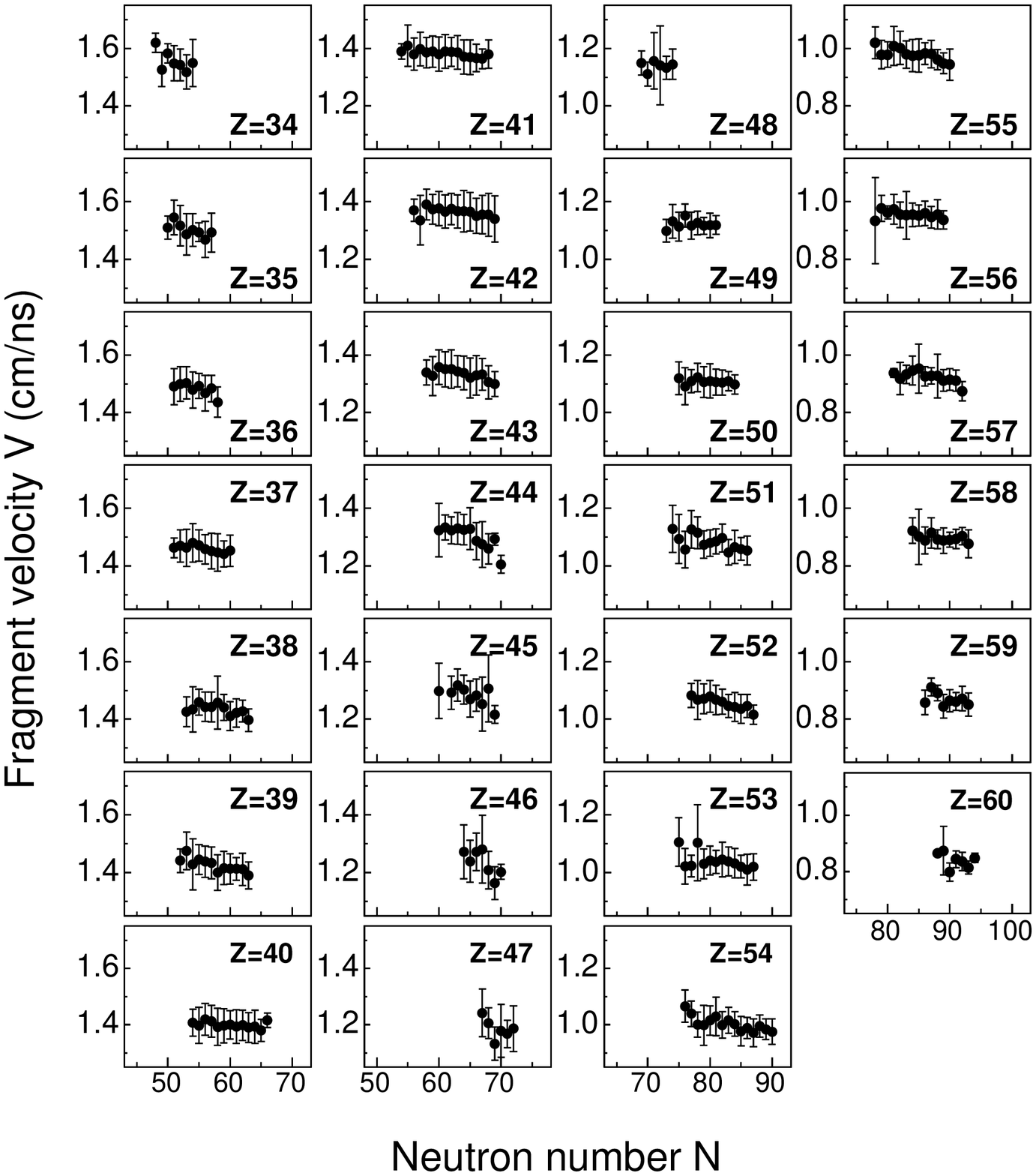}%
\includegraphics[width=\columnwidth]{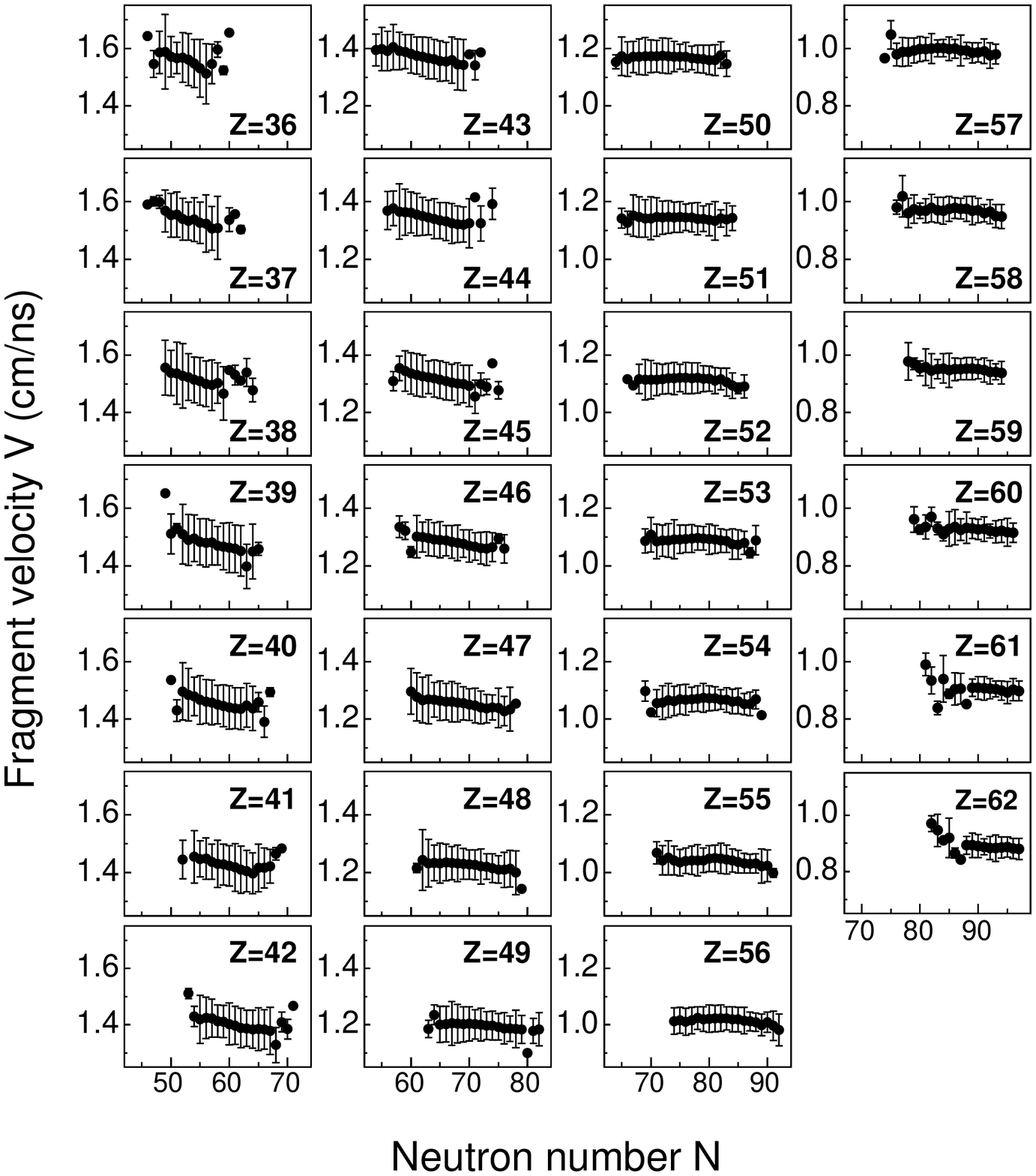}
\caption{Mean values of the fission fragment velocity spectra as a function of the neutron number of the elements produced in the fission of $^{240}$Pu (left panels) and $^{250}$Cf (right panels). The error bars show the second momentum of the velocity spectra.}
\label{vel_iso_cf}
\end{center}
\end{figure*}

In the present paper, the kinematic properties of the fission fragments are used to trace back information on the original split at scission. Due to the advantages of inverse kinematics, the average neutron excess of the fragments can be derived experimentally for the first time, in addition to the total kinetic energy and the neutron multiplicities. These results put forward unique information on the resulting equilibrium of neutron and proton sharing in the elongation process up to the scission point. The distance between the fragments and their total excitation energy at scission are also determined and discussed.

\section{Fission velocities}

The characteristics of the fragment distributions from fission of two different compound nuclei, $^{240}$Pu and $^{250}$Cf, have been recently investigated~\cite{caaPRC2013}. Both fissioning systems were produced using a $^{238}$U beam impinging onto a $^{12}$C target at 6.1~AMeV, an energy 10\% above the Coulomb barrier. The $^{240}$Pu fissioning system was populated in ($^{12}$C,$^{10}$Be) transfer reactions. The detection of the target-like nuclei allowed the selection of the different reaction channels and, with the measurement of the energy and angle of the target-like nuclei, the excitation energy produced in the transfer reaction is reconstructed event by event~\cite{xavier}. In this work, it is assumed that all the excitation energy produced in the reaction is carried by the beam-like fissioning system. This approximation has been validated in a recent investigation of the transfer process between $^{238}$U and $^{12}$C~\cite{carme}. Following this assumption, $^{240}$Pu was produced with an excitation energy distribution centered on $E^*\sim 9$~MeV and a width of 6~MeV. In the case of $^{250}$Cf, the fissioning system was the result of fusion reactions between the $^{238}$U beam and the $^{12}$C target. The detection of the target-like nuclei allowed the selection of the different reaction channels. The VAMOS spectrometer~\cite{Savajols19991027c, Pullanhiotan} was used for the identification of the fission fragments and permitted a measure of the complete isotopic distribution of the fission fragments for these well-defined fissioning systems.

\begin{figure*}[!t]
\begin{center}
\includegraphics[width=\columnwidth]{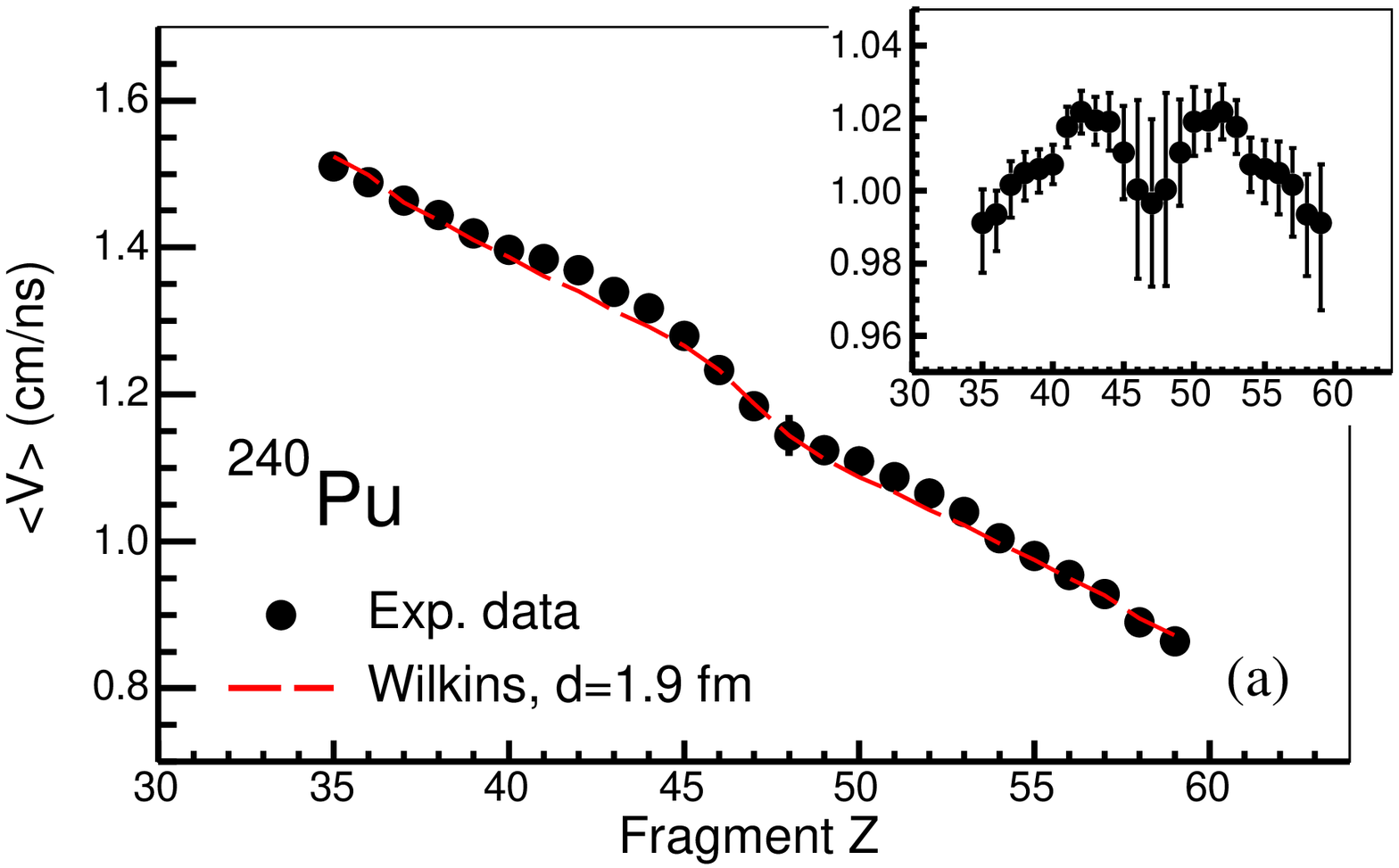}%
\includegraphics[width=\columnwidth]{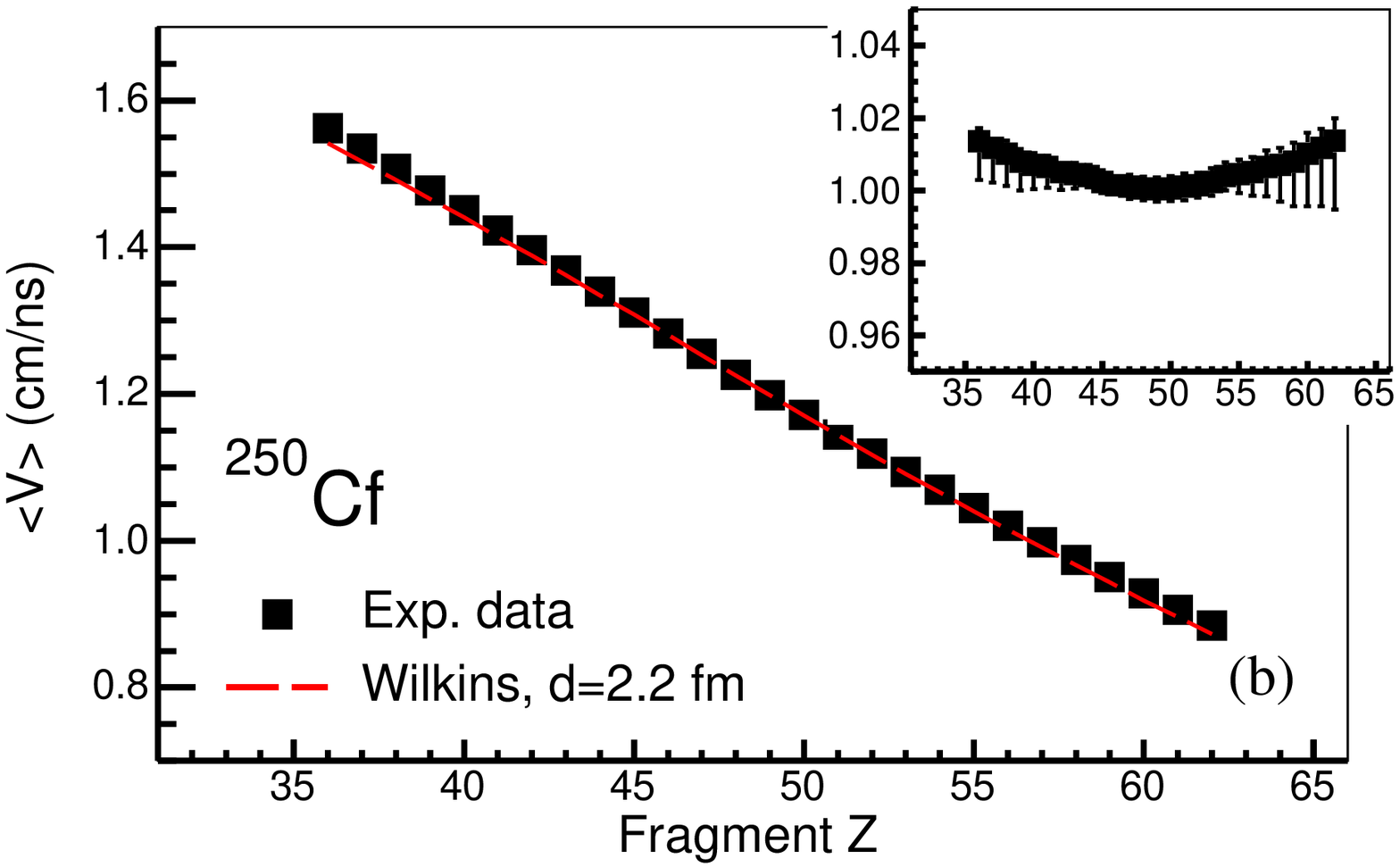}
\caption{(Color online) Average fragment velocity $\langle V\rangle$ as a function of the fragment atomic number $Z$ produced in the fission of $^{240}$Pu (a) and $^{250}$Cf (b). Red lines follow the prescription of Wilkins {\it et al.}~\cite{wilkins} with constant deformation and neck parameters, and the masses deduced in Sec.~\ref{sec_mass} (see text for details). The insets show the ratio between the measured data and the calculations.}
\label{vel_Z_pu}
\end{center}
\end{figure*}

At the focal plane of the spectrometer, the trajectory of the transmitted fission fragments was determined using two position measurements, giving access to the focal plane vertical and horizontal angles. The measurement of the time of flight and a trajectory-reconstruction procedure, based on the diagonalization of the spectrometer transport matrix~\cite{Pullanhiotan}, were used to reconstruct the fragment velocity vectors. From this knowledge and the information on the kinematics of the compound nucleus, it is possible to reconstruct the fission-fragment velocity vectors in the reference frame of the fissioning nucleus, which defines the fission velocity vector. In the present work, the fission velocity determination is improved with respect to the previous work~\cite{caaPRC2013}: the slowing down of the fission fragments in the target has been taken into account, whereas it was considered as negligible previously. For each fragment, the velocity measured in the laboratory is then corrected for the energy loss following the prescription of Ref.~\cite{dufour}, in which the different parameters are adjusted by means of LISE++ simulations~\cite{LISE}. In addition, the velocity distributions of each fragment have been corrected for transmission cuts (angle and ionic charge states) that modified the mean value of the velocity distribution by few percents. The resolution of the resulting fission velocities in the reference frame of the fissioning system depends on the resolution of the velocity and angle in the laboratory reference frame, the beam-energy straggling, and the angle and energy of the target-like product in the case of transfer-induced fission. Considering a resolution of 0.4\% on the velocity measurement and an angular resolution of 5 mrad~\cite{NIMB266}, the resolution on the resulting fission velocity is better than 2\%. At the limits of the VAMOS acceptance, heavier fragments are less transmitted and mass distributions suffer from cuts in their heavier part. These cuts also impact the velocity determination. The effect of the acceptance on the velocity, which can modify the value of the measurement in $\sim 0.5$\%, is included in the error bars. The resulting velocity distributions in the reference frame of the fissioning nucleus $V(A,Z)$ from fission of $^{240}$Pu and $^{250}$Cf are displayed in Fig.~\ref{vel_iso_cf}.

 The average velocity $\langle V\rangle$ for each atomic number $Z$ is calculated with the fragment yields $Y(Z,A)$ as:
 \begin{equation}
 \langle V\rangle (Z)=\frac{\sum\limits_A V(Z,A)\cdot Y(Z,A)}{\sum\limits_A Y(Z,A)},
 \label{eq_v_ave}
 \end{equation}
 
 They are displayed in Fig.~\ref{vel_Z_pu} for both systems. The steady decrease with increasing $Z$, observed for both cases, reveals the momentum conservation that induces smaller velocities as the mass of the fragment increases. In general, the velocities measured in fission from $^{250}$Cf are between 2\% and 10\% larger than those of $^{240}$Pu due in part to the stronger Coulomb repulsion between the fragments, which are heavier in the case of $^{250}$Cf. The average velocity $\langle V\rangle$ is compared with the prescription of Wilkins {\it et al.}~\cite{wilkins} (see Eqs.~\ref{eq_tke} and \ref{eqD}), with constant deformation and neck parameters. In this formulation, the mass associated to each fragment atomic number was from Eq.~\ref{eqa*}, as explained in the following section. In both fissioning systems, the deformation parameters were kept at a constant value of $\beta_1=\beta_2=0.625$~\cite{sch00} for the two fragments while the neck distance is set in order to match the values of $\langle V\rangle$ at symmetric splits. In the case of $^{240}$Pu, the neck is then $d=1.9$~fm long, while for $^{250}$Cf it needs to be increased to $d=2.2$~fm. The inset of Fig.~\ref{vel_Z_pu}(a) shows clear deviations of the measured velocities from fission of $^{240}$Pu with respect to those calculated with fixed deformation and neck parameters around $Z\sim52$ and $Z\sim42$. These are the signature of smaller deformation in the scission configuration appearing for specific proton and neutron numbers, as will be discussed later. In the case of $^{250}$Cf, the deviations show that for asymmetric splits, the measured velocities are larger than those calculated with fixed $\beta_1=\beta_2=0.625$ and $d=2.2$~fm.

\section{Reconstruction of the fragment mass at scission}
\label{sec_mass}
The fission-fragment velocities contain information on important properties of the scission configuration such as deformation and masses of the nascent fragments. In the reference frame of the fissioning system and due to momentum conservation, the ratio of the two fragment velocities is equal to the inverse ratio of the initial masses $M^*_{1,2}$ on each fission event:
\begin{equation}
\frac{V_1\gamma_1}{V_2\gamma_2}=\frac{M^*_2}{M^*_1},
\label{eqva}
\end{equation}
with $\gamma=[1-(V/c)^2]^{-1/2}$.

\begin{figure*}[!t]
\begin{center}
\includegraphics[width=\columnwidth]{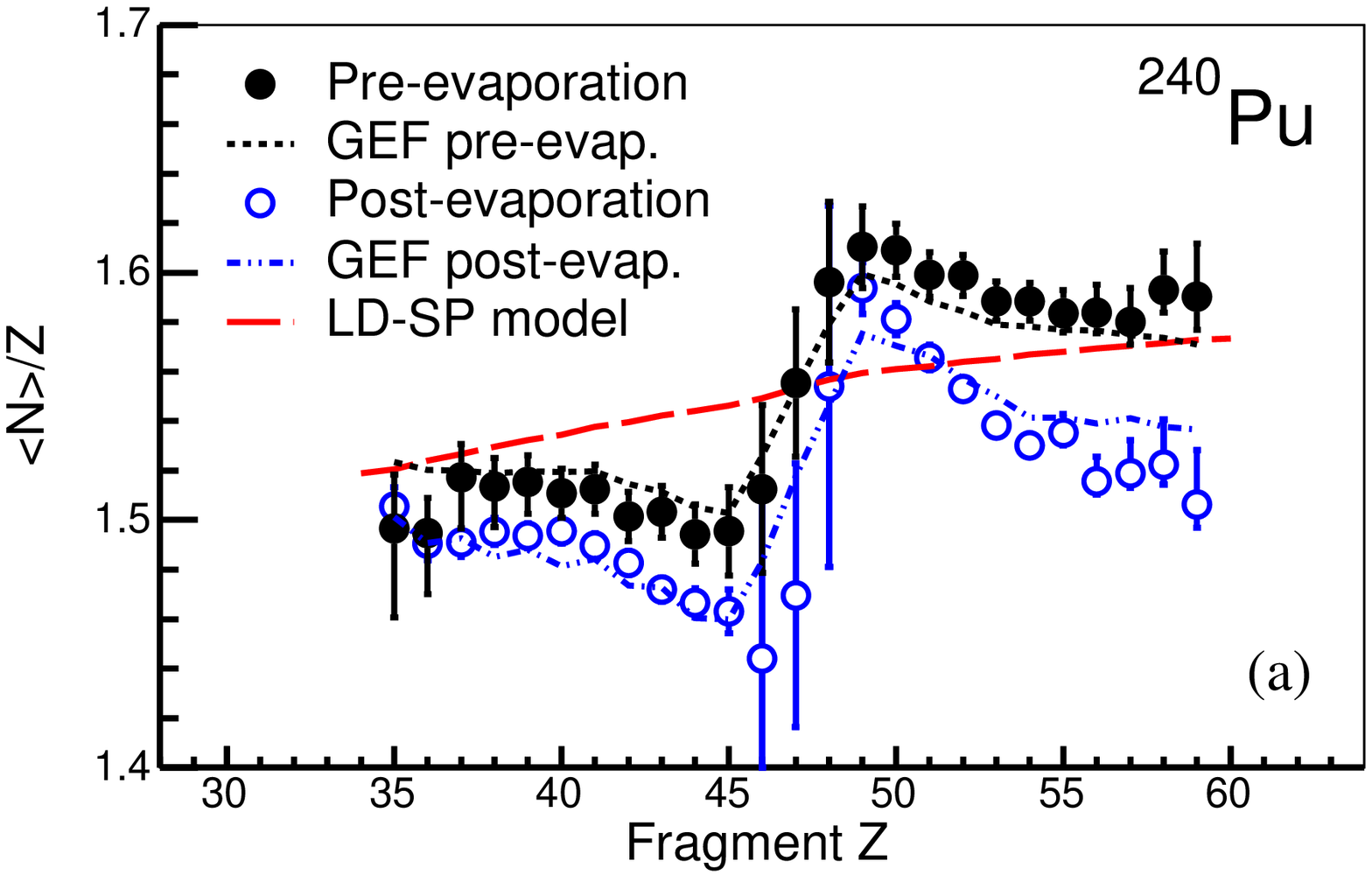}%
\includegraphics[width=\columnwidth]{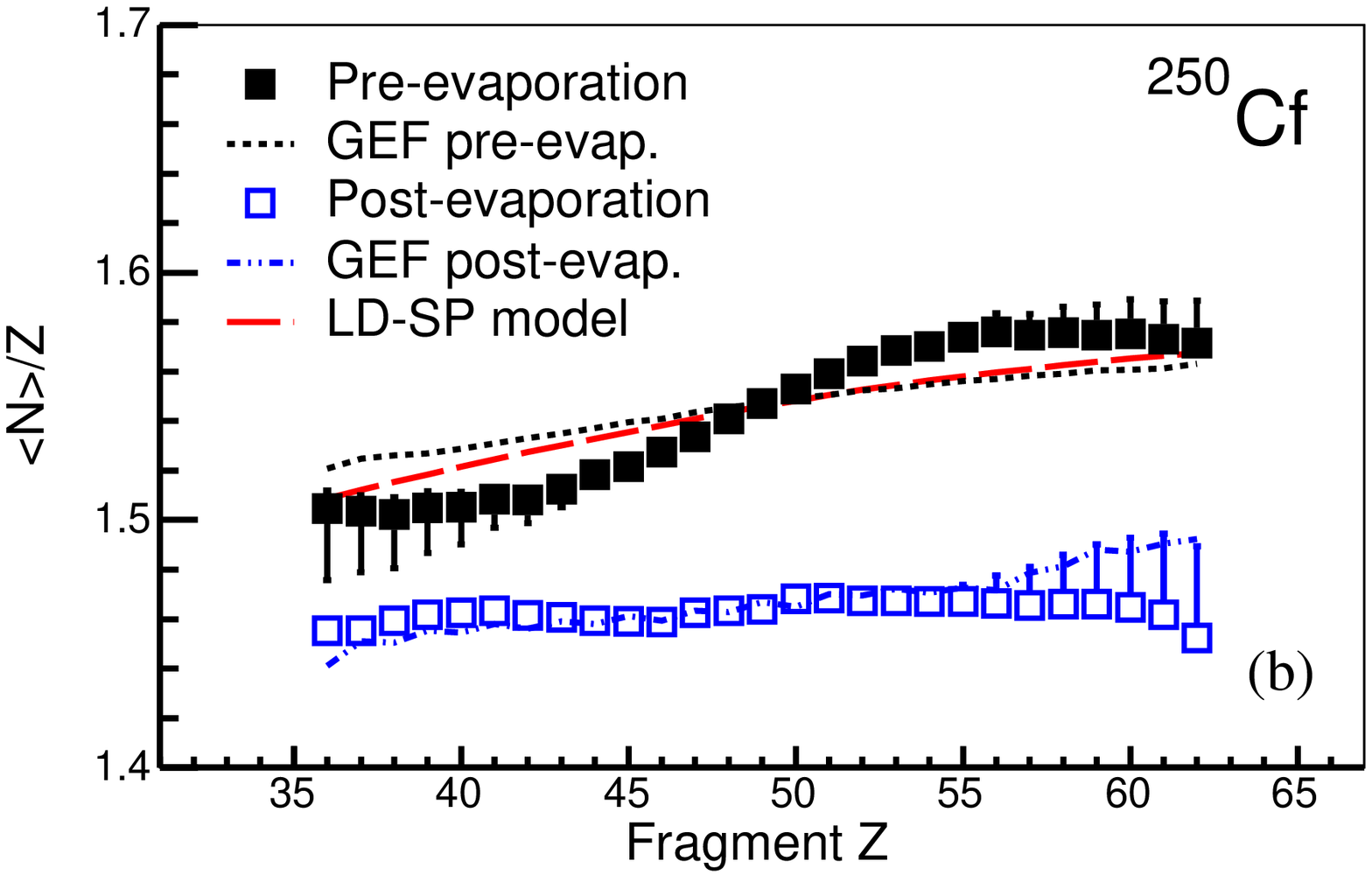}
\caption{(Color online) The average neutron excess $\langle N\rangle/Z$ of fragments produced in the fission of $^{240}$Pu (a) and  $^{250}$Cf (b) as a function of their atomic number $Z$, is displayed before (black markers) and after (blue open markers) neutron evaporation. It is compared to the estimation of the GEF code before (black dashed line) and after (blue dashed-dotted line) evaporation. Red long-dashed lines correspond to the predictions of the LD-SP model.}
\label{NZ_pre}
\end{center}
\end{figure*}

In the present experiment, the velocity of one fragment per fission event is measured. This is done over the complete fragment production, as shown in the preceding section. In both investigated cases, the excitation energy of the fissioning systems is not enough to allow proton evaporation along the fission process, therefore, for a fissioning system of atomic number $Z_{\mathrm{FS}}$, the average characteristics, including momentum, of fragments with atomic number $Z_1$ are associated to the complementary atomic number $Z_2=Z_{\mathrm{FS}}-Z_1$. Under these conditions, the momentum conservation can relate both fragments as:
\begin{equation}
\langle V_1\gamma_1 M^*_1\rangle =\langle V_2\gamma_2 M^*_2\rangle,
\end{equation}
where the observables with subscript 1 correspond to the fragments with atomic number $Z_1$ and those with subscript 2 correspond to its complementary fragment $Z_2=Z_{\mathrm{FS}}-Z_1$. 
The analysis of the measured fragments shows that $\langle V_i\gamma_i M^*_i\rangle$ can be approximated to $\langle V_i\gamma_i\rangle\langle M^*_i\rangle$ with a relative error smaller than 0.4\%. The average values of mass excess permit to express $M^*_i$ as the product of the number of nucleons and the unified atomic mass unit, $M^*_i\approx u A^*_i$, with a deviation smaller than 0.01\% in $A^*_i$. In addition, a detailed simulation of the isotropic neutron evaporation by fragments, within ranges of velocity, mass, and $Z$ as the ones measured, confirms that its effect on the fragment velocity is smaller than 0.15\%. These approximations allows the use of the relation of Eq.~\ref{eqva} with the measured velocities averaged for each fragment:
\begin{equation}
\frac{\langle V_1\gamma_1\rangle}{\langle V_2\gamma_2\rangle}=\frac{\langle A^*_2\rangle}{\langle A^*_1\rangle}.
\label{eqva2}
\end{equation}
Assuming that no neutron evaporation occurs from saddle to scission nor from the neck at scission, and knowing the mass of the fissioning system $A_{\mathrm{FS}}$, the average masses at scission can be deduced:
\begin{equation}
\begin{array}{cc}
\langle A_1^*\rangle=&A_{\mathrm{FS}}\frac{\langle V_1\gamma_1\rangle}{\langle V_1\gamma_1\rangle+\langle V_2\gamma_2\rangle},\\
\langle A_2^*\rangle=&A_{\mathrm{FS}}-\langle A_1^*\rangle.
\label{eqa*}
\end{array}
\end{equation}

In the case of $^{240}$Pu fissioning system, with an average excitation energy of 9~MeV, no pre-scission neutron evaporation is considered; however, a contamination of 20\% from $^{241}$Pu \cite{caaPRC2013} renders the final fissioning mass $A_{\mathrm{FS}}=240.2$ for transfer-induced fission. In the case of $^{250}$Cf, the excitation energy is sufficient to produce neutron evaporation before fission with a multiplicity that depends on the excitation energy and angular momentum induced in the reaction. In a similar reaction with $^{16}$O beam on $^{238}$U, angular anisotropy has been used to determine the root-mean-squared angular momentum induced in the reaction at an energy close to the Coulomb barrier~\cite{backPRC32}. A value of 24$\hbar$ was determined for this reaction. Calculations based on the Bass modelisation~\cite{gontcharCPC107}, adjusted to reproduce the angular reaction of $^{16}$O$+^{238}$U data, give an estimation of 20$\hbar$ in the case of the present experiment. This angular momentum and an initial excitation energy of 45~MeV lead to an average fissioning system of mass $A_{\mathrm{FS}}=249.6$, considering the different fission probabilities obtained from GEF~\cite{GEF} predictions, where first-chance fission happens in more than 68\% of the events. This result is confirmed by the measurement of pre-scission neutrons emitted from a slightly lighter compound nuclei with similar excitation energy and larger angular momentum, for which a multiplicity of $\sim 1$ is found~\cite{sax94}. The uncertainty estimated on the fissioning-nucleus average mass is restricted to 0.3 mass units, considering possible small variations on angular momentum and the validity of the model. The pre-scission neutron evaporation also modifies the average excitation energy of the fissioning system, estimated with the same code in $E^*\sim 42$~MeV.

The experimental neutron excess of the fission fragments is known to vary with the asymmetry of the fission split~\cite{boc89}. In the framework of unchanged charge density, both fragments keep the same neutron excess as the fissioning nucleus. Experimentally, it has been observed that the light fragments are less neutron rich than this prediction, while the heavy fragments are more neutron rich. This is known as the charge polarization. The neutron excess at scission $\langle N^* \rangle/Z$ is calculated with the average masses $\langle A^* \rangle$ obtained in Eq.~\ref{eqa*} as:
\begin{equation}
\langle N^*\rangle/Z (Z)=\frac{\langle A^* \rangle(Z)-Z}{Z},
\end{equation}
 and it is shown in Fig.~\ref{NZ_pre} with full symbols as a function of the fragment $Z$ for $^{240}$Pu and $^{250}$Cf. The deduced $\langle N^* \rangle/Z$ at scission is compared in this figure with the measured $\langle N \rangle/Z$ after neutron evaporation, shown in open symbols. In this case, the average $\langle N \rangle$ is obtained from the measured masses and fragment yields~\cite{caaPRC2013} as:
\begin{equation}
\langle N \rangle (Z)=\frac{\sum\limits_A A\cdot Y(Z,A)}{\sum\limits_A Y(Z,A)}-Z.
\end{equation}

The error bars in Fig.~\ref{NZ_pre} include the statistical uncertainty, the estimated deviation due to acceptance cuts, the evaluated uncertainty of the fission velocity in the reference frame of the fissioning system, and of the mass of the fissioning system. In the case of post-neutron $\langle N \rangle/Z$ for $^{240}$Pu, the large error bars around symmetry are produced by the combined effect of low yields and contamination from other channels. It is also important to note the different nature of the calculation of both post- and pre-neutron neutron excess: the former is evaluated with the direct identification in ($Z$, $A$) of the fission fragments, while the latter is derived from a continuous observable such as the velocity.

\begin{figure*}[!t]
\begin{center}
\includegraphics[width=\columnwidth]{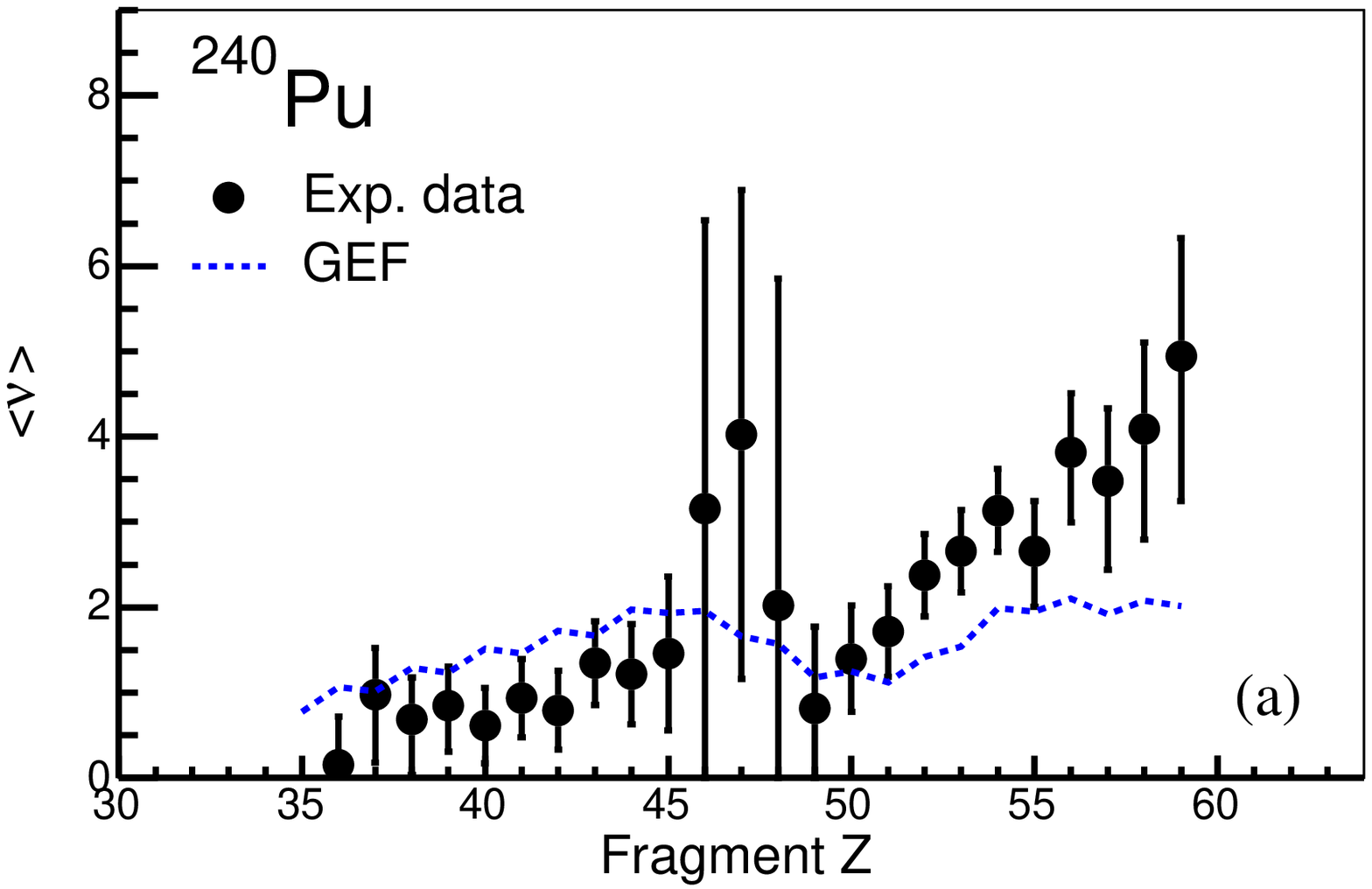}%
\includegraphics[width=\columnwidth]{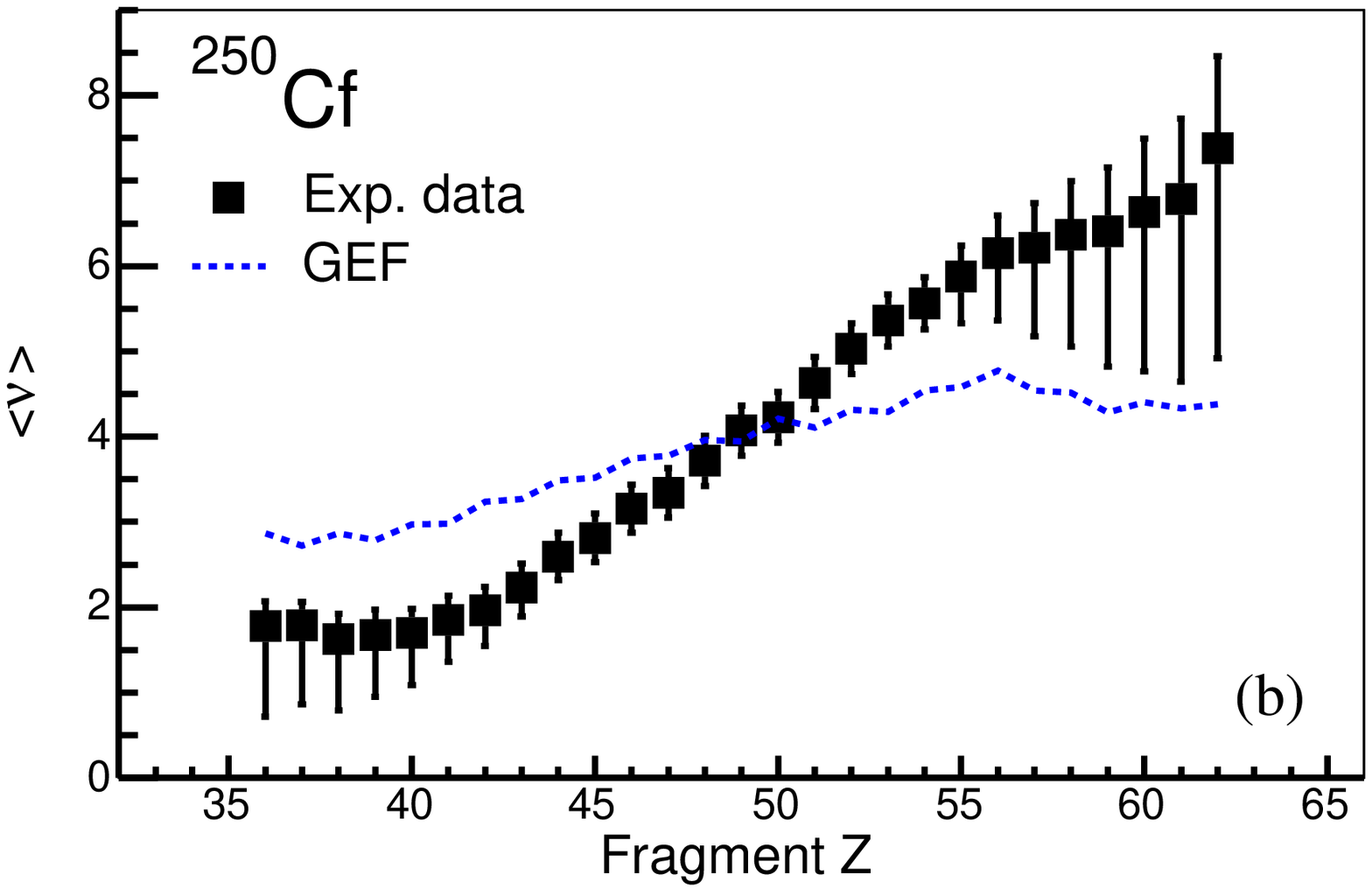}\\
\includegraphics[width=\columnwidth]{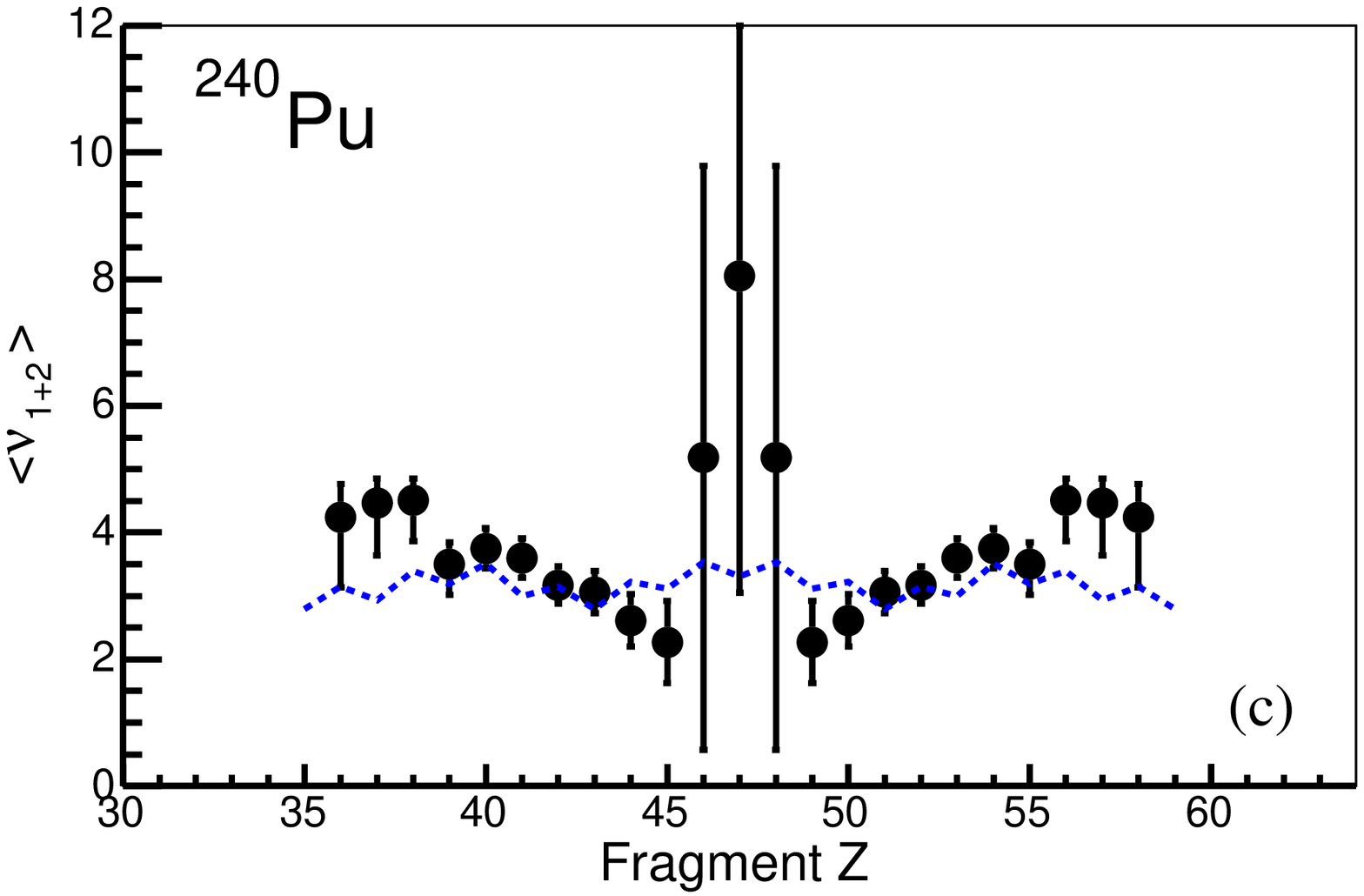}%
\includegraphics[width=\columnwidth]{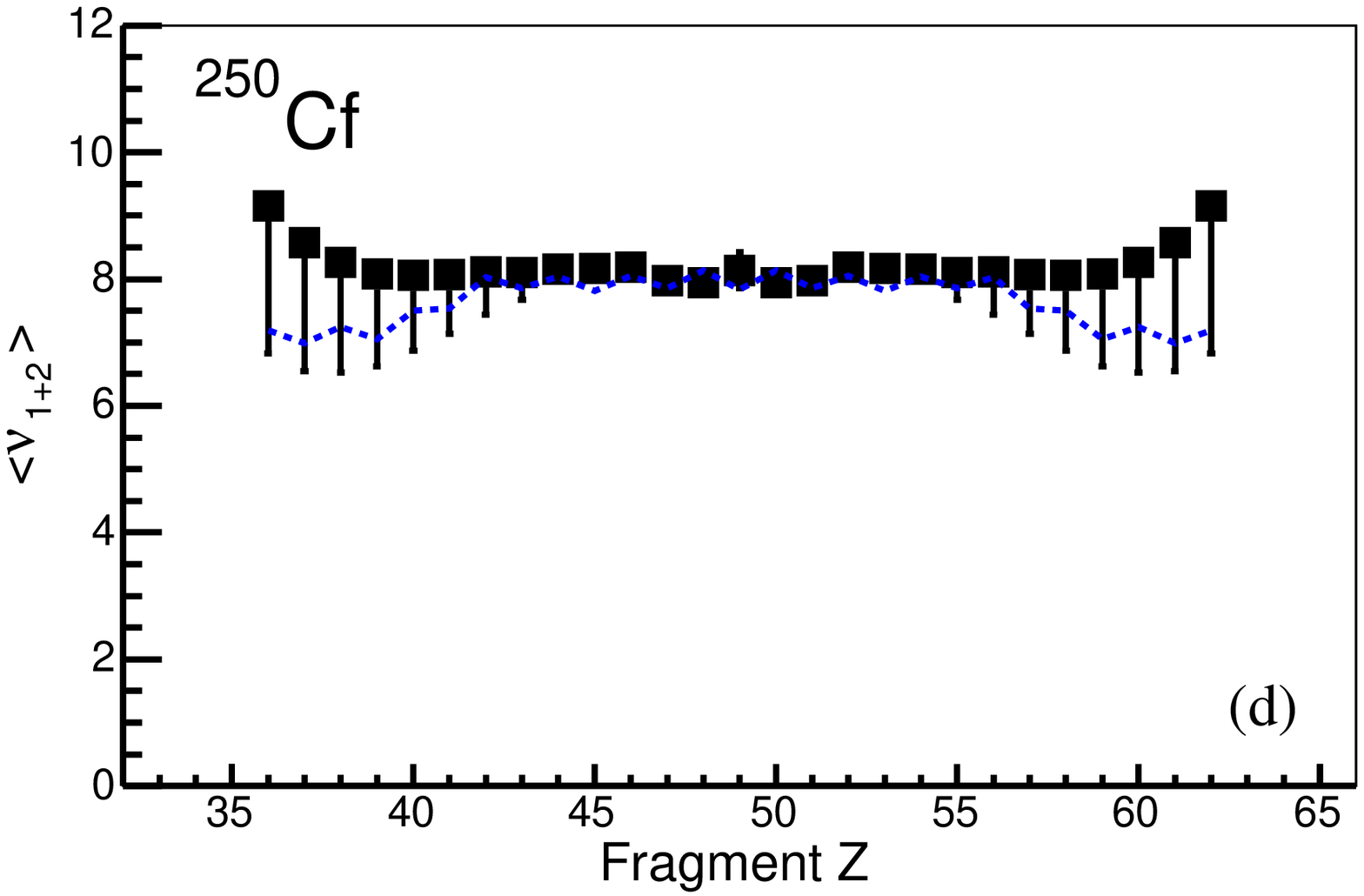}
\caption{(Color online) Average neutron multiplicity $\langle\nu \rangle$ (a,b) and total neutron evaporation multiplicity $\langle\nu_{1+2}\rangle$ (c,d) for $^{240}$Pu and $^{250}$Cf. Blue dotted lines correspond to the calculations performed with the GEF code.}
\label{nu}
\end{center}
\end{figure*}

The resulting $\langle N^*\rangle/Z$ for fission of $^{240}$Pu shows a step behaviour with a sudden increase that peaks around $Z\sim 50$, paired with a minimum around $Z\sim 44$. The subsequent neutron evaporation does not attenuate the step structure, which is also observed in $\langle N\rangle/Z$ (open symbols in Fig.~\ref{NZ_pre}). In the case of fission of $^{250}$Cf, $\langle N^*\rangle/Z$ shows a smooth behaviour and a steady increase with $Z$, while neutron evaporation modifies significantly the trend. The fragments observed after neutron evaporation show a rather constant neutron excess, with no charge polarization. The smaller error bars observed in the fragments of $^{250}$Cf reflect the larger statistics in the population of this fissioning system compared to that of $^{240}$Pu, due to the cross sections of fusion-fission reactions being about a factor 50 larger than that of transfer reactions~\cite{carme, biswasPRC56}.

The deduced $\langle N^*\rangle/Z$ can be compared with a simple calculation based on the scission-point model introduced by Wilkins {\it et al.}~\cite{wilkins}, where the most probable combination of neutron excess of both fragments is defined as the one that minimizes the total energy of the system. In the model used in the present work, only macroscopic energies of the deformed fragments are considered, following the liquid-drop energy prescription of Myers and Swiatecki~\cite{myers}, including Coulomb interaction and neglecting nuclear interaction between the two touching fragments. The results of this liquid-drop scission-point (LD-SP) calculation are displayed in Fig.~\ref{NZ_pre}. They show a steady increase of the neutron excess of scission fragments with their atomic number, reflecting the need to increase the number of neutrons for counterbalancing the increasing Coulomb repulsion inside the heavier fragments. This behavior is observed in experimental data from $^{250}$Cf, where the compound nucleus fissions at $E^*\sim 42$~MeV of excitation energy. Still, the experimental neutron excess deviates from this smooth behaviour with a sort of ``S'' shape that renders the heavier fragments slightly more neutron rich, and thus lighter fragments less neutron rich, than expected from equilibration of macroscopic strengths. Stronger deviations are observed in the case of $^{240}$Pu, which is produced with an average excitation energy of 9~MeV, due to the influence of the nuclear structure on the sharing of neutrons and protons between the fragments. Apart from these deviations, the deduced $\langle N^*\rangle/Z$ and the calculation of the LD-SP model coincide for very asymmetric splits in both systems.

A more realistic calculation, including the effect of nuclear structure and neutron evaporation, can be done with the GEF code. Figure~\ref{NZ_pre} shows the neutron excess calculated with GEF at scission and after neutron evaporation for $^{240}$Pu and $^{250}$Cf respectively. The structures appearing in the neutron excess of the $^{240}$Pu scission and final fragments are well reproduced by GEF, with a certain underestimation of the neutron excess for heavier fragments and therefore an overestimation for lighter ones. These structures are attributed to the effect of shells of protons and/or neutrons that tend to minimize the system energy around certain nucleon numbers in the nascent fragments, modifying their production when compared to pure macroscopic (or liquid-drop) behavior. At higher excitation energy, these effects are expected to be very much reduced~\cite{Ignatyuk} and the calculations with the GEF code are consistent with this expectation. However, a clear deviation of the experimental data with respect to the code appears in the case of $^{250}$Cf fission at $E^*\sim 42$~MeV: while the calculations with GEF show a steady increase of the neutron excess with the $Z$, as observed in the data, a significant deviation of this slope is found between $Z\sim 44$ and $Z\sim 54$, rendering light fragments less neutron rich and heavy ones more neutron rich than those obtained with the code. Despite this difference, the GEF code reproduces the experimental constant value of the measured $\langle N\rangle/Z$ of the final fragments as a function of their atomic number. The features of $\langle N^*\rangle/Z$ and $\langle N\rangle/Z$ suggest the need to improve the description of the potential at scission, as well as the sharing of excitation energy, that modifies the neutron excess of fragments between scission and final fragments. 



\section{Post-scission neutron evaporation}
\label{sec_nevap}
The difference between the scission mass deduced from the velocity properties and the measured mass at the focal plane of the spectrometer gives access to the average neutron multiplicity $\langle\nu\rangle$ evaporated by the deformed scission fragments:
\begin{equation}
\langle\nu\rangle (Z)=\langle A^*\rangle (Z)-\langle A\rangle(Z).
\end{equation}

The resulting average neutron multiplicities are shown in Fig.~\ref{nu}. In the case of $^{240}$Pu, a saw-tooth behaviour of the neutron multiplicity, usually observed as a function of mass~\cite{tsuchiya}, can be recognized despite the large fluctuations that appear in the tails of the function and for symmetric splits (see Sec.~\ref{sec_mass}). The saw-tooth shape of the neutron multiplicity in low-energy fission is understood as the result of the deformation energy released when the scission fragments separate and recover their ground-state deformation~\cite{nishio}. For higher-energy fission, the additional excitation energy is shared among the two fragments, while part of it might be evaporated during the saddle-to-scission path. This additional energy will be also released in part with evaporated neutrons, thus making the neutron multiplicity a probe of the energy sharing between the two fragments. In order to reproduce the observed multiplicities, models describing the available phase space in each fragment on the level density of a Fermi gas need to introduce additional parameters~\cite{talou}. A more accurate reproduction of the measured multiplicities is achieved with descriptions based on the evolution of the level density as the excitation energy increases from constant-temperature to Fermi-gas models~\cite{esorting}. The neutron multiplicities obtained with the GEF code, shown in Fig.~\ref{nu}, are calculated according this last description. The comparison with the data shows a similar behaviour, although with an overestimation of $\sim 1$ neutron for light fragments and a similar underestimation for heavier fragments. These differences between the data and the code are related with those of $\langle N^*\rangle/Z$ and $\langle N\rangle/Z$ previously described. 
The experimental multiplicity $\langle\nu \rangle$ of fission fragments of $^{250}$Cf at $E^*\sim 42$~MeV is displayed in Fig.~\ref{nu}(b). A steady increase with $Z$ is observed, in agreement with direct measurements of neutron multiplicity in similar fusion-fission reactions~\cite{hinde92,kozulin}. The kink observed at $Z\sim 54$ can also be seen in data from (p+$^{242}$Pu) around $A\sim 140$~\cite{kozulin}. GEF calculations show a similar increase, though with a less steep slope. This difference may arise from the treatment on the evolution from constant-density to a Fermi-gas description and/or the description of the polarization at scission.

\begin{figure*}[!t]
\begin{center}
\includegraphics[width=\columnwidth]{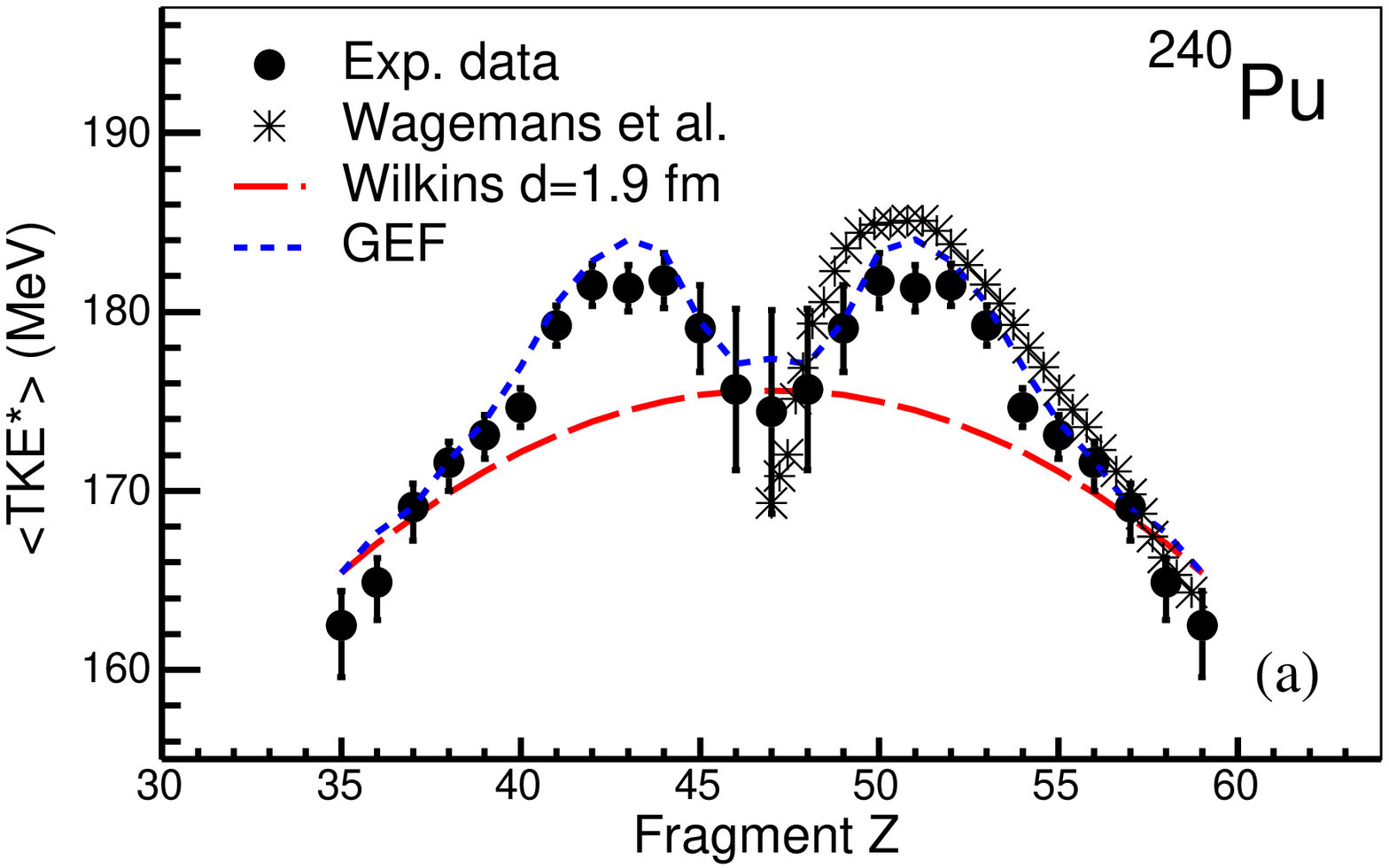}%
\includegraphics[width=\columnwidth]{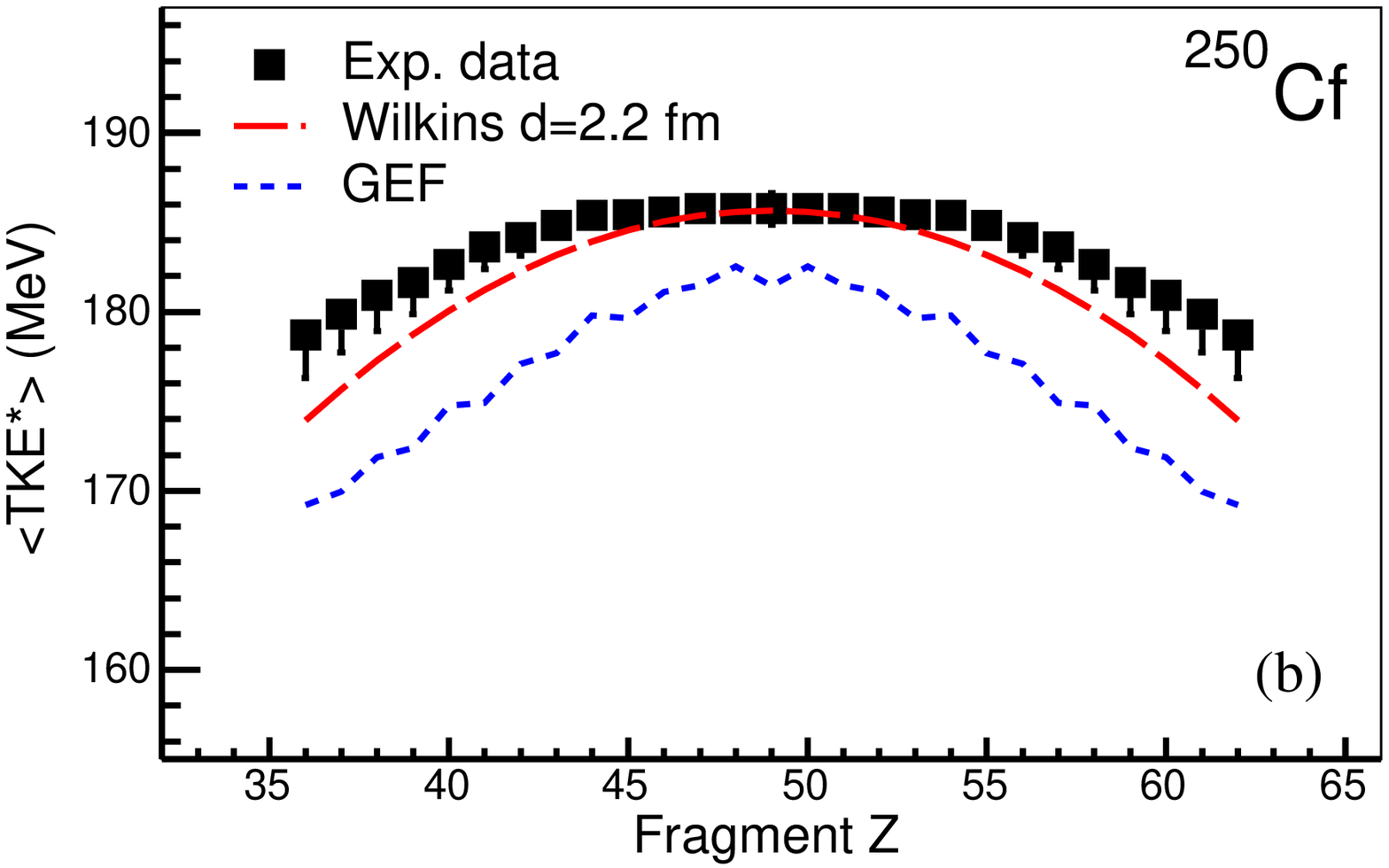}
\caption{(Color online) Total kinetic energy of fission fragments at scission of $^{240}$Pu (a) and $^{250}$Cf (b). Red dashed lines correspond to the prescription of Wilkins {\it et al.} (see Eq.~\ref{eq_tke}), displayed for reference, with constant deformation parameters $\beta_1=\beta_2=0.625$, and $d=1.9$~fm for $^{240}$Pu and $d=2.2$~fm for $^{250}$Cf. Blue dashed lines correspond to the calculations of the GEF model. Asterisks show data from thermal-neutron induced fission of $^{239}$Pu~\cite{wag84}.}
\label{tke}
\end{center}
\end{figure*}

It is worth noting that the comparison of the calculations performed with GEF with the experimental data is more sensitive when studying the average multiplicity as a function of the fragment $\langle\nu\rangle$ than in the case of the post-scission total multiplicity per fission, defined as the sum of the multiplicities of fragment $Z_1$ and its complementary $Z_2=Z_{\mathrm{FS}}-Z_1$, or the mass difference between the fissioning system and the fragments detected after neutron evaporation:
\begin{equation}
  \langle\nu_{1+2}\rangle = \langle\nu_1\rangle + \langle\nu_2\rangle = A_{\mathrm{FS}} - \langle A_1\rangle - \langle A_2\rangle. 
\end{equation}

The total multiplicity $\langle\nu_{1+2}\rangle$ from fragments of $^{240}$Pu and $^{250}$Cf are shown in Figs.~\ref{nu}(c,d). The total multiplicity of $^{240}$Pu shows an average value around $\langle\nu_{1+2}\rangle\approx 3$ and an increase for symmetric and very asymmetric splits, where the deformation of fragments is expected to be maximal. However, as previously noted, the region around symmetry for this system suffers from large uncertainties. In the case of $^{250}$Cf, a flat behaviour without structure is observed. The different features revealed in the study of $\langle \nu\rangle$ are hidden in the case of the total multiplicity $\langle\nu_{1+2}\rangle$. In Fig.~\ref{nu}, $\langle\nu_{1+2}\rangle$ is compared with calculations performed with the GEF code. As in the case of $\langle\nu\rangle$, the total multiplicity $\langle\nu_{1+2}\rangle$ of $^{240}$Pu is well reproduced, within the large statistical fluctuations. In the case of $^{250}$Cf, a similar constant behaviour around symmetry is observed in the GEF calculations.

\section{Total kinetic energy at scission}

The characteristics of the total kinetic energy $TKE$ distribution of fully accelerated fragments are mainly determined by the configuration of the system at scission and also, when measured in most experiments, by the following neutron evaporation. By using the fragment masses reconstructed at scission and their velocities, it is possible to derive the Coulomb potential energy that acts between the fragments that would be transformed in $TKE$ once the fragments are fully accelerated. In order to simplify the nomenclature, this potential energy will be referred as average Total Kinetic Energy at scission $\langle TKE^*\rangle$, and it is the equivalent to the total kinetic energy after full acceleration but before prompt-neutron emission. This quantity keeps a similar information on the scission configuration as the $TKE$ but without the influence of neutron evaporation. The average $\langle TKE^*\rangle$ is calculated as:
\begin{equation}
\langle TKE^*\rangle=u\langle A^*_1\rangle\left(\langle\gamma_1\rangle-1\right) + u\langle A^*_2\rangle\left(\langle\gamma_2\rangle-1\right),
\end{equation}
with $u=931.494$~MeV. The resulting $\langle TKE^*\rangle$ is displayed in Fig.~\ref{tke} for both fissioning systems. Ignoring the contribution of pre-scission velocity, the total kinetic energy at scission can be considered as the result of the Coulomb repulsion between the fragments at a distance $D$~\cite{wilkins}:
\begin{equation}
TKE^*=1.44\frac{Z_1Z_2}{D}.
\label{eq_tke}
\end{equation}
the distance between the two charged centroids $D$ can be expressed as~\cite{wilkins}:
\begin{equation}
D=r_0{A_1^*}^{1/3}\left(1+\frac{2}{3}\beta_1\right) + r_0{A_2^*}^{1/3}\left(1+\frac{2}{3}\beta_2\right)+d,
\label{eqD}
\end{equation} 
where $r_0=1.16$~fm, $\beta_1$ and $\beta_2$ are the quadrupole deformation parameters of fragments of mass number $A_1^*$ and $A_2^*$, and $d$ is the distance between the surface of both fragments, associated with the length of the neck. Figure~\ref{tke} shows a comparison between the experimental $\langle TKE^*\rangle$ and this expression with constant parameters of deformation $\beta_1=\beta_2=0.625$, and a neck parameter of $d=1.9$~fm in the case of $^{240}$Pu and $d=2.2$~fm in the case of $^{250}$Cf. In both cases, the neck parameter was chosen to match the experimental values at symmetric splits, where the condition $\beta_1=\beta_2$ is trivially fulfilled and the fragment mass at scission is $A_{\mathrm{FS}}/2$. The masses of the fragments are those reconstructed at scission, as explained in Sec.~\ref{sec_mass}. The comparison with fission from $^{240}$Pu shows strong deviations around $Z\sim 50$ and $Z\sim 44$. These values of $\langle TKE^*\rangle$, higher by about 10~MeV, correspond to more compact shapes, and are usually interpreted as the result of the superposition of different fission modes corresponding to the formation of fragments with atomic spherical and deformed closed shells~\cite{boeck08}. Figure~\ref{tke}(a) shows the same features in thermal-neutron induced fission of $^{239}$Pu~\cite{wag84} and a good agreement with calculations performed with the GEF code. 

In the case of $^{250}$Cf, the measured $\langle TKE^*\rangle$ produces an almost flat behaviour around symmetric splits, between $Z\sim 44$ and $Z\sim 54$, while the difference with respect to the calculation with fixed deformation and neck parameters increases up to $\sim 4$~MeV towards asymmetric splits. The GEF code shows an overall underestimation from about $\sim 4$~MeV at symmetry to $\sim 10$~MeV for asymmetric splits, while it follows a similar shape as that with fixed scission parameters. In order to calculate $\langle TKE^*\rangle$, GEF uses different approaches for fission with low- and high-excitation energy. In the former case, $\langle TKE^*\rangle$ is calculated as the residue of the total energy after subtracting the collective, intrinsic, and deformation energies. In the case of high excitation energy, GEF uses a similar approach as the LD-SP model, where the energy at scission is minimized to find the corresponding $\beta_1$ and $\beta_2$ deformation parameters, while keeping an empirical value of $d=3$~fm for the neck. In addition, a certain amount of velocity pre-scission is added, and the influence of fragment angular momenta is also considered. 

The relation between the elongation, namely, the distance between fragments, and the resulting $\langle TKE^*\rangle$ can be seen and quantified more clearly considering the ratio of the actual distance $D$ between the fragments at scission and their distance $D_0$ as spherical touching nuclei. Following the prescription of Eq.~\ref{eq_tke}, the average of this ratio $\langle D/D_0\rangle$ as a function of the fragment split can be calculated as:
\begin{equation}
\label{eq_DD0}
\langle D/D_0\rangle=1.44\frac{Z_1Z_2}{\langle TKE^*\rangle}\left(\frac{1}{r_0{\langle A_1^*\rangle}^{1/3} + r_0{\langle A_2^*\rangle}^{1/3}}\right).
\end{equation}

\begin{figure}[!t]
\begin{center}
\includegraphics[width=\columnwidth]{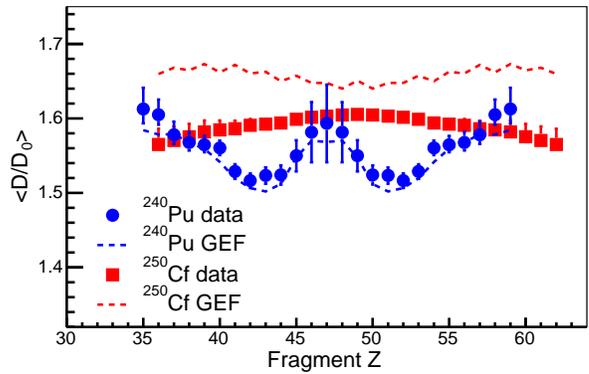}
\caption{(Color online) Ratio between the distance between fragments at scission and that as spherical touching nuclei for fission of $^{240}$Pu (blue dots) and $^{250}$Cf (red squares). Dashed lines correspond to calculations done with the GEF code for $^{240}$Pu (blue dashed line) and $^{250}$Cf (red dashed line).}
\label{dd0}
\end{center}
\end{figure}

Figure~\ref{dd0} shows $\langle D/D_0\rangle$ for fission of both $^{240}$Pu and $^{250}$Cf. The most compact shapes appear around $Z\sim 44$ and $Z\sim 50$, as observed in the $\langle TKE^*\rangle$, while at symmetric and very asymmetric splits, the elongation reaches values 60\% larger than the distance between the corresponding spherical touching fragments. In the case of $^{250}$Cf, the elongation is kept around a value of $\langle D/D_0\rangle\sim1.60$, with a slight increase at the symmetry and no apparent effects attributed to nuclear structure. Figures~\ref{tke} and~\ref{dd0} show that the elongation of the system at scission changes with the asymmetry of the split, modifying the experimental $\langle TKE^*\rangle$. Also in Fig.~\ref{dd0}, the $\langle D/D_0\rangle$ calculated with the GEF code is shown. The comparison with the $^{240}$Pu data is consistent with the agreement found in both the neutron excess and the $\langle TKE^*\rangle$. In the case of $^{250}$Cf, GEF predicts overall more elongated configurations with an almost constant value of $D/D_0\sim 1.66$ and a slight dependence on $Z$ somehow opposite to the one found in the experimental data: in the case of GEF, symmetric splits are less elongated than asymmetric ones. Together with the situation found in the $\langle TKE^*\rangle$, this behaviour suggests a more complex dependence on the fragment split of both the deformation and neck distance at scission with high excitation energies and relatively high angular momentum. 

\begin{figure*}[!t]
\begin{center}
\includegraphics[width=\columnwidth]{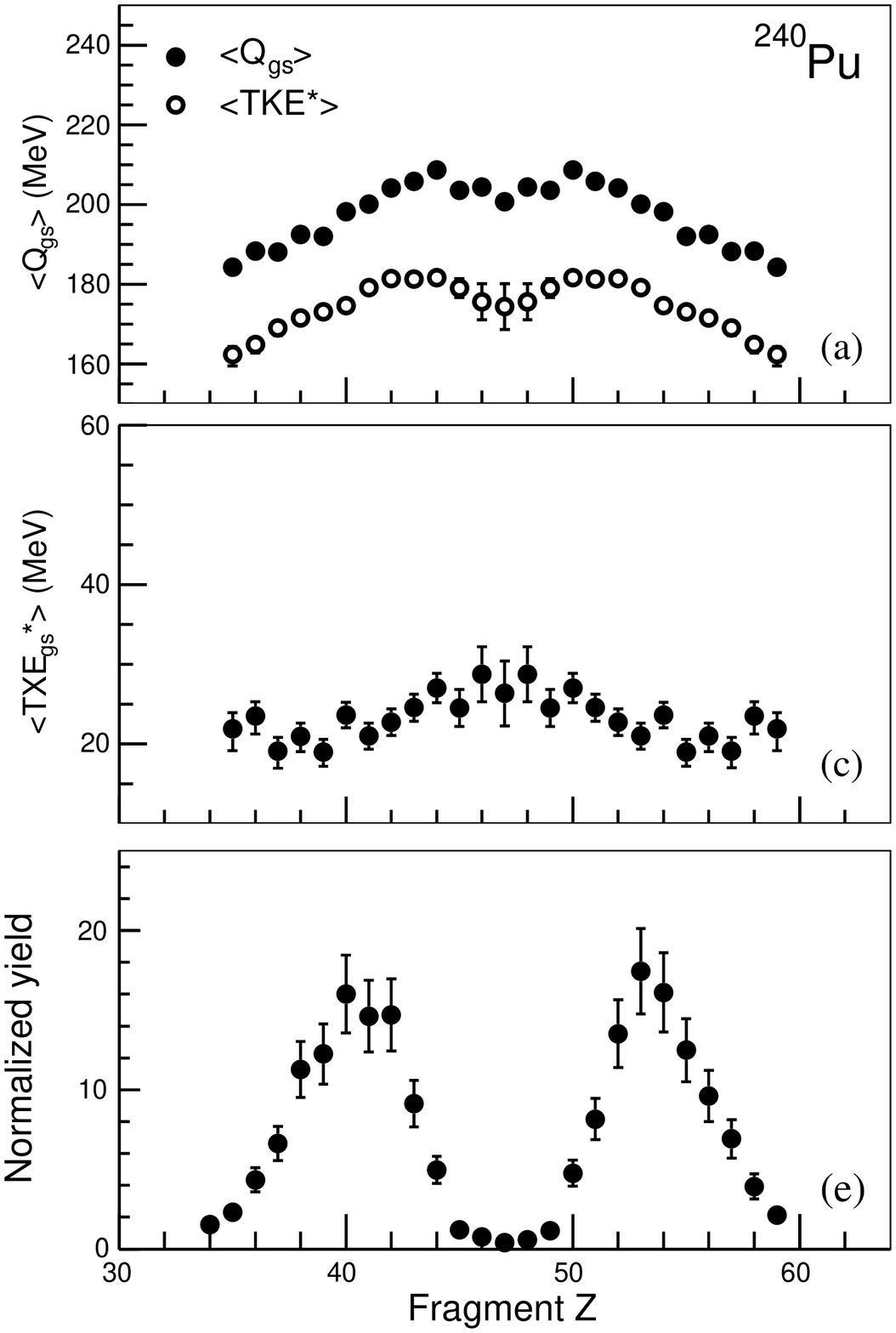}
\includegraphics[width=\columnwidth]{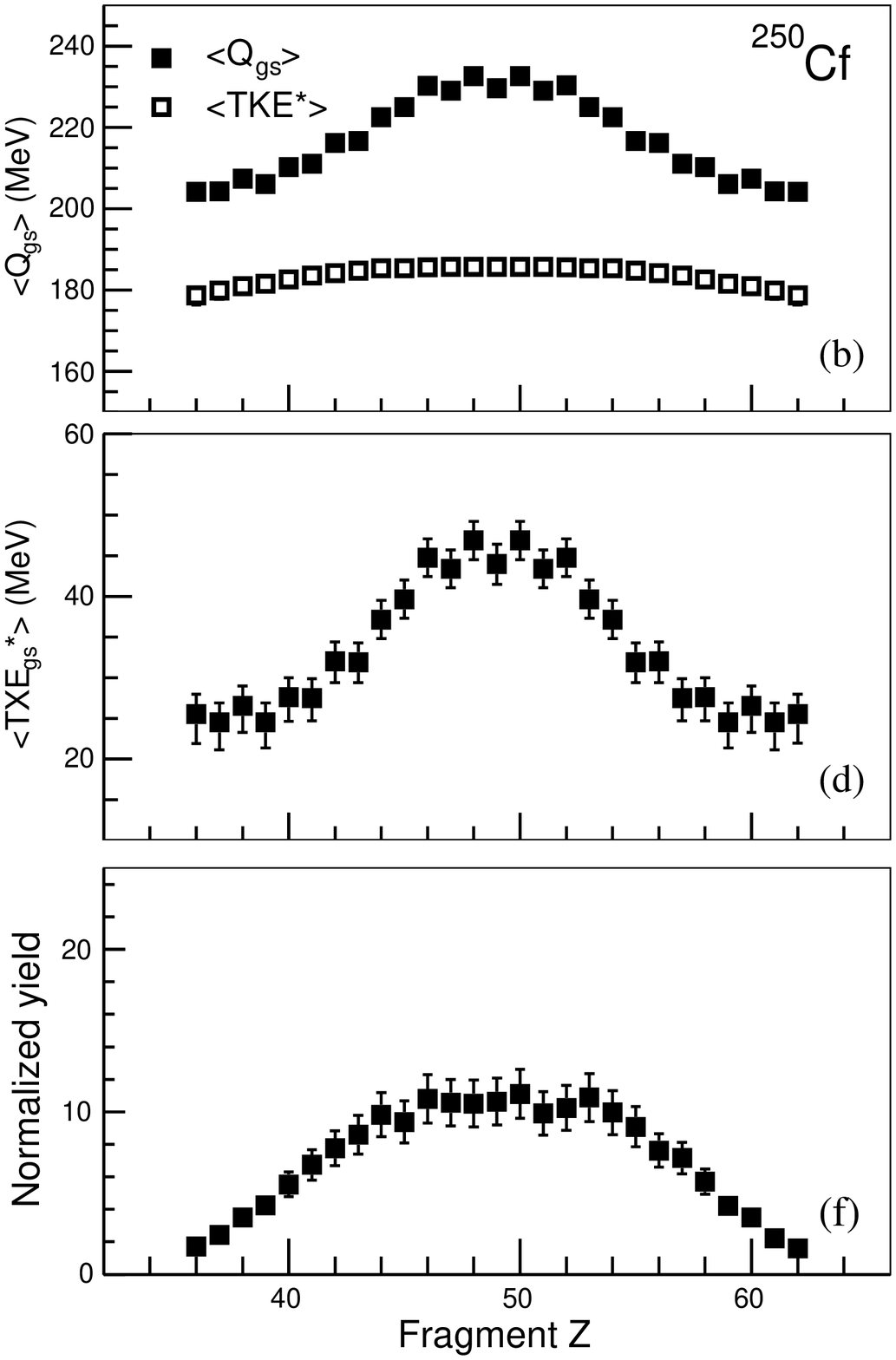}
\caption{(a,b) Average difference between the ground states of the fissioning system and the fragments $\langle Q_{\mathrm{gs}}\rangle$ at scission (full symbols) and the $\langle TKE^*\rangle$ (open symbols). (c,d) Total excitation energy at scission $\langle TXE^*_{\mathrm{gs}}\rangle$. (e,f) Post-neutron evaporation normalized yields~\cite{caaPRC2013} for reference. (a), (c), and (e) correspond to $^{240}$Pu while (b), (d), and (f) correspond to $^{250}$Cf.}
\label{potsurf}
\end{center}
\end{figure*}

\section{Total excitation energy at scission}
The total energy available in the fission reaction $E_{\mathrm{tot}}$ is the sum of the mass of the fissioning system $M_{\mathrm{FS}}$ in its ground state and its excitation energy $E^*_{\mathrm{FS}}$. At scission, $E_{\mathrm{tot}}$ is the sum of the masses of the fragments, the $TKE^*$, and the excitation of collective (deformation, rotation, etc.) and intrinsic (single-particle excitations) degrees of freedom. The experimental determination of $\langle TKE^*\rangle$ at scission allows then to determine the portion of the total energy that is transformed in the excitation energy of the fragments. Ignoring neutron evaporation during the saddle to scission path, the conservation of $E_{\mathrm{tot}}$ links both instances:
\begin{equation}
\begin{array}{cl}
E_{\mathrm{tot}}&=M_{\mathrm{FS}}+E^*_{\mathrm{FS}}\\
~&=\langle M^*_1\rangle + \langle M^*_2\rangle + \langle TKE^*\rangle + \langle TXE^*\rangle,\\
\end{array}
\end{equation}
where $\langle M^*_i\rangle$ is the average ground-state mass of the fragment $i$ and $\langle TXE^* \rangle$ is the total excitation energy available for both fragments. The calculation of $\langle M^*_i\rangle$ from the obtained $\langle A^*_i\rangle$ is done by interpolation of the mass of the closest integer mass numbers:
\begin{equation}
\begin{array}{rl}
\langle M^*_i\rangle=&(1-W_i)\cdot m_0\left(Z_i,\lfloor\langle A^*_i\rangle\rfloor\right)+\\
 ~&+ W_i\cdot m_0\left(Z_i,\lfloor\langle A^*_i\rangle\rfloor+1\right),\\
W_i=&\langle A^*_i\rangle-\lfloor\langle A^*_i\rangle\rfloor,
\end{array}
\label{eq_M}
\end{equation}
where $m_0\left(Z_i,A_i\right)$ is the ground-state mass of the isotope $(Z_i,A_i)$ from the AME2012 evaluation~\cite{audi}. The floor operator $\lfloor\langle A^*_i\rangle\rfloor$ gives the largest integer not greater than $\langle A^*_i\rangle$. The quantity $\langle TXE^* \rangle$ can also be regarded as the difference between the total energy $E_{\mathrm{tot}}$ and the the potential surface at scission. By subtracting $E^*_{\mathrm{FS}}$, the depth of the potential can be expressed with respect to the ground state of the fissioning system as:
\begin{equation}
\begin{array}{cl}
\langle TXE^*_{\mathrm{gs}}\rangle &=M_{\mathrm{FS}}-\langle M^*_1\rangle-\langle M^*_2\rangle-\langle TKE^*\rangle\\
~&=\langle Q_{\mathrm{gs}}\rangle-\langle TKE^*\rangle,
\end{array}
\label{eq_txe}
\end{equation}
\begin{figure*}[!t]
\begin{center}
\includegraphics[width=\columnwidth]{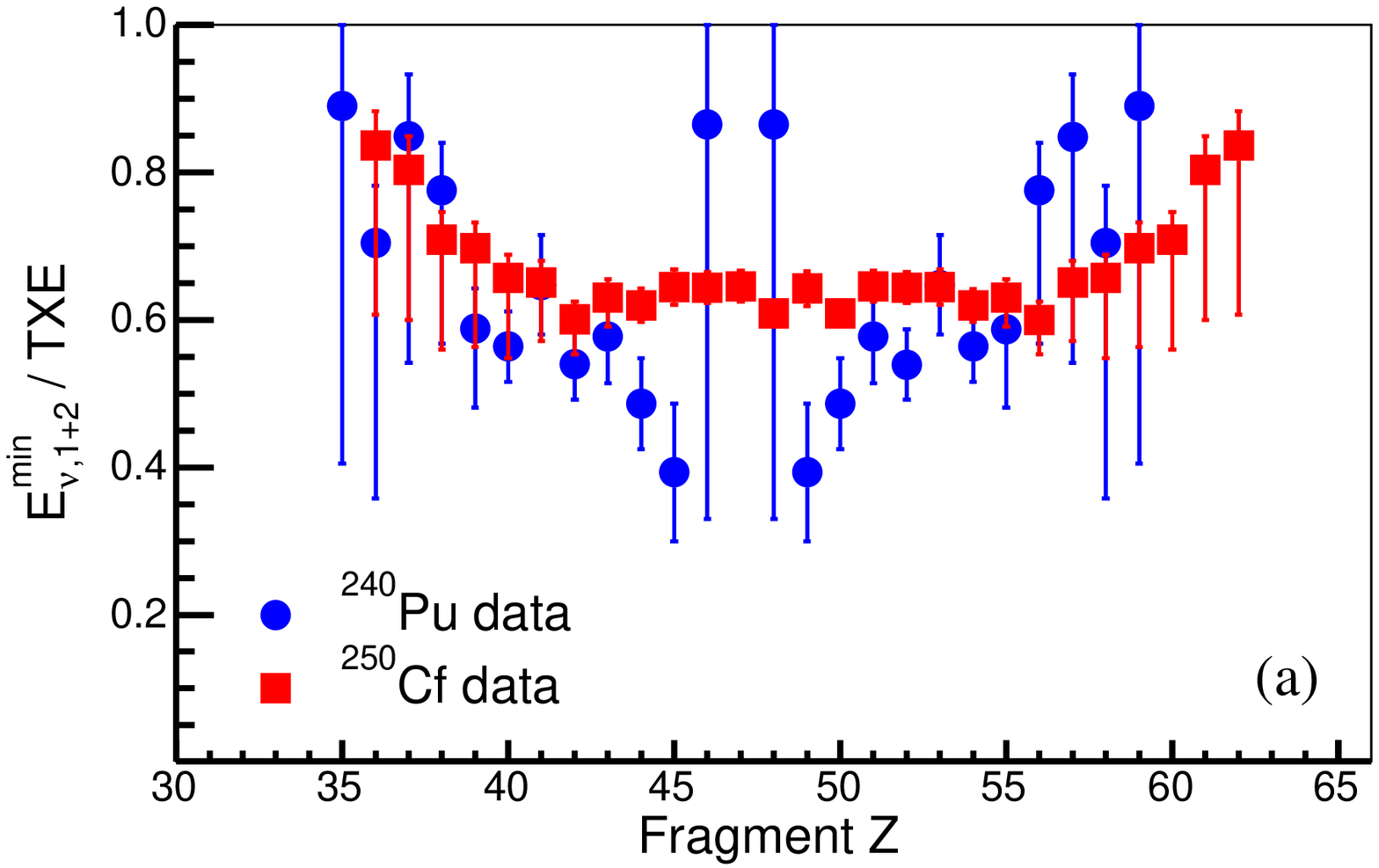}%
\includegraphics[width=\columnwidth]{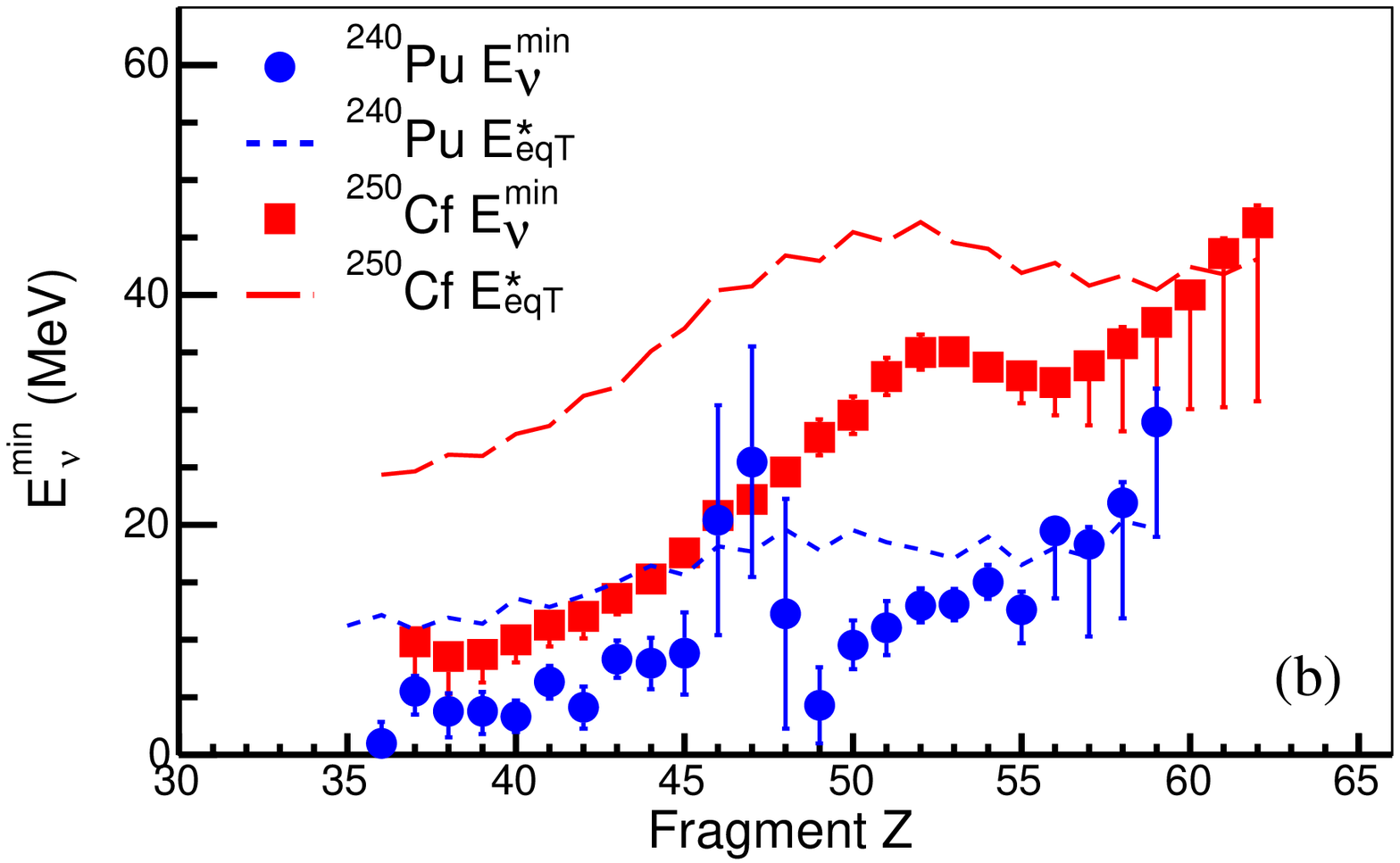}%
\caption{(Color online) (a) Minimum fraction of the total excitation energy $\langle TXE^*\rangle $ evaporated with neutrons for $^{240}$Pu (blue dots) and $^{250}$Cf (red squares). (b) Minimum energy released in neutron evaporation and excitation energy shared according with the mass ratio of the fragments for $^{240}$Pu (blue dots and dashed line) and $^{250}$Cf (red squares and long-dashed line).}
\label{fEnmtot}
\end{center}
\end{figure*}
with $\langle Q_{\mathrm{gs}}\rangle$, displayed in Figs.~\ref{potsurf}(a,b), as the average mass difference between the fissioning system and the fragments at scission. The error bars include the estimated deviation of the masses $\langle M^*_i\rangle$ calculated from $\langle A^*_i\rangle$ with Eq.~\ref{eq_M} and those calculated from actual distributions of individual $M^*(Z,A)$. This deviation is found to be below 2~MeV. Both systems exhibit a $\langle Q_{\mathrm{gs}}\rangle$ with strong even-odd staggering, due to the systematic production of odd-odd or even-even $Z$ pairs of fragments. The behaviour of $\langle Q_{\mathrm{gs}}\rangle$ for each system is similar to that of their respective $\langle TKE^* \rangle$ (see Fig.~\ref{tke}), with two maxima around $Z\sim 44$ and $Z\sim 50$ in the case of $^{240}$Pu and a flatter top in the case of $^{250}$Cf. Figures~\ref{potsurf}(c,d) show the quantity $\langle TXE^*_{\mathrm{gs}}\rangle$ as a function of $Z$ for $^{240}$Pu and $^{250}$Cf. Both cases have a similar behaviour with a well-defined maximum centered at symmetric splits that raises from $\sim 20$ up to $\sim 25$~MeV in $^{240}$Pu and from $\sim 25$ up to $\sim 45$~MeV in $^{250}$Cf. In Ref.~\cite{asghar}, a liquid-drop model (LDM) that considers surface, surface asymmetry, and Coulomb contributions is applied to thermal-neutron induced fission of $^{239}$Pu, resulting in a $TXE_{\mathrm{gs}}^*=15.27$~MeV compared to $\sim 25$~MeV determined in the present work. The same calculation applied to spontaneous fission of $^{252}$Cf results in $TXE_{\mathrm{gs}}^*=24.94$~MeV, much smaller than the $\sim 45$~MeV found for $^{250}$Cf. The main differences between the systems used in the LDM calculations and those of this work correspond to the excitation energies and angular momenta produced in the reactions inducing fission. In the case of $^{240}$Pu, thermal-neutron induced reactions yield an excitation energy of $E^*=6.53$~MeV and negligible angular momentum, while the transfer-induced fission of this work produces a distribution of excitation energy centered at $E^*\sim 9$~MeV and a distribution of angular momentum of the order of $L\sim 6\hbar$ (see~\cite{gava} for a similar reaction). The spontaneous fission of $^{252}$Cf induces neither excitation energy or angular momentum, while the fusion-induced fission measured in this work produces $^{250}$Cf with $E^*\sim 42$~MeV and an average $L\sim 20\hbar$.

The presented values of $\langle D/D_0\rangle $ and $\langle TXE_\mathrm{gs}^*\rangle $ as functions of $Z$ (Figs.~\ref{dd0} and \ref{potsurf}) can be interpreted as the shape of the potential landscape that the fissioning system experiences at scission. In both systems, symmetric splits are characterized by elongated shapes and maxima of potential. In the case of $^{240}$Pu, signatures of the most produced asymmetric yields, around $Z\sim 42$ and $Z\sim 52$, are well correlated with shorter distances between the fragments and features in $\langle TKE^*\rangle $ around $Z\sim 44$ and $Z\sim 50$ (see Fig.~\ref{tke}(a)), that can be interpreted as the effect of nuclear structure. However, the deduced $\langle TXE_\mathrm{gs}^*\rangle $ for the same system does not show any distinct behaviour around these atomic numbers. The maxima in $\langle TKE^*\rangle$ and $\langle Q_\mathrm{gs}\rangle$ compensate when both are subtracted to calculate $\langle TXE_\mathrm{gs}^*\rangle $, leaving no clear signature in the depth of the potential at scission.

The total excitation energy gained at scission is released by the fragments via post-scission neutron and gamma evaporation until their ground state is reached. The energy needed to evaporate these neutrons is the sum of their binding energies and their kinetic energy, which is a distribution between 0 and some MeV (see~\cite{hinde92,nishio} for instance). Therefore, the sum of the separation energies of the evaporated neutrons is a lower limit for the energy released by neutron evaporation. From the total neutron multiplicity $\langle\nu_{1+2}\rangle$ (see Sec.~\ref{sec_nevap}) and the post- and pre-evaporation masses, this limit for the average energy released by neutron evaporation $E_{\nu,1+2}^\mathrm{min}$ can be estimated as a function of the fragment split as:
\begin{equation}
E_{\nu,1+2}^\mathrm{min}=\langle M_1\rangle+\langle M_2\rangle+m_n\cdot\langle\nu_{1+2}\rangle-\langle M_1^*\rangle-\langle M_2^*\rangle,
\label{eq_enmintot}
\end{equation}
where $\langle M_i\rangle$ and $\langle M_i^*\rangle$ are the ground-state masses post- and pre-evaporation of fragment $i$, and $m_n$ is the neutron mass. The neutron multiplicity evaporated after scission contains information on the excitation energy, including deformation energy, stored by the fragments at scission. In order to access this information from multiplicity measurements, is important to know the fraction of excitation energy carried by neutron evaporation. Figure~\ref{fEnmtot}(a) shows the ratio between $E_{\nu,1+2}^\mathrm{min}$ and the deduced $\langle TXE^*\rangle $ for the two systems studied. The behaviour of $E_{\nu,1+2}^\mathrm{min}$ is similar to the ones observed in the neutron multiplicity in Fig.~\ref{nu}, as expected. In both systems, the fraction of $\langle TXE^*\rangle $ evaporated with neutrons is smaller at symmetry, with $^{240}$Pu reaching some 50\% and $^{250}$Cf showing a flat behaviour between $Z\sim 44$ and $\sim 54$ with more than 60\% of $\langle TXE^*\rangle $. The two systems take more $\langle TXE^*\rangle $ to evaporate neutrons, reaching up to 80\%, as the split becomes more asymmetric, although a flatter behaviour would be within the error bars.

The minimum energy released in neutron evaporation can be also calculated as a function of the fragment $Z$:
\begin{equation}
E_{\nu}^\mathrm{min}(Z)=\langle M\rangle(Z)+m_n\cdot\langle\nu\rangle(Z)-\langle M^*\rangle(Z).
\label{eq_enmin}
\end{equation}
Figure~\ref{fEnmtot}(b) shows $E_{\nu}^\mathrm{min}$ as a function of $Z$ for both systems. As in the case of $E_{\nu,1+2}^\mathrm{min}$, the behaviour is similar to the ones observed in the neutron multiplicity in Fig.~\ref{nu}, with the heavy fragments, particularly in $^{250}$Cf, releasing more excitation energy in the form of neutron evaporation. This energy $E_{\nu}^\mathrm{min}$ is a fraction of the excitation energy that each fragment stores up to scission, the rest being released as gamma emission. For fission with high $E^*_\mathrm{FS}$, where structure effects vanish, it is expected that the total $TXE$ is shared between the fragments according with their mass ratio at scission~\cite{toke}:
\begin{equation}
E_{\mathrm{eqT}}^*(Z)=\langle TXE^*\rangle\frac{\langle A^*\rangle(Z)}{A_{\mathrm{FS}}},
\label{eq_Exeq}
\end{equation}
being $E_{\mathrm{eqT}}^*$ the excitation energy of the fragment, following this prescription. It is important to note that, in any case, structure effects would not modify the $E^*$ partition for symmetric or very asymmetric splits (see~\cite{chen11}, for instance). For lower energy fission, this behaviour is less valid, even at very asymmetric splits, for reasons different from pure structure effects. Direct measurements of neutron multiplicity show that the heavy fragment keeps more excitation energy than Eq.~\ref{eq_Exeq} or structure effects predict~\cite{naqvi}. The evolution of the level density and the temperature with the mass of the fragment prevents them to reach equal temperatures, with the heavier fragment always cooler than the lighter one, and thus dragging more energy, resulting in an excitation energy of the hevier fragments larger than the predicted by Eq.~\ref{eq_Exeq}~\cite{esorting}. Figure~\ref{fEnmtot}(b) compares the deduced $E_{\nu}^\mathrm{min}$ with $E_\mathrm{eqT}^*$. In both cases, the $E_{\nu}^\mathrm{min}$ of heavier fragments reaches similar values as their corresponding $E_\mathrm{eqT}^*$, while light fragments take less than the half of it. This behaviour suggests that, in both systems, the actual $E^*$ of the heavy fragments should be larger than $E_\mathrm{eqT}^*$ in order to allow for the kinetic energy of the neutrons and gamma evaporation. This observation favors the description of a continuous flow of $E^*$ from the light to the heavy fragments due to their unbalanced temperatures.

\section{Conclusions}

With the new generation of high-quality fission data obtained in inverse kinematics, new observables of the fission process are available. From the reconstructed fission kinematics, the total kinetic energy $\langle TKE^*\rangle$ and average masses of the two fragments can be reconstructed at scission as a function of the fragment atomic number. In this work, the neutron excess of fragments is determined at scission for two different fissioning systems, $^{240}$Pu and $^{250}$Cf, with different excitation energies. For the first time, an experimental insight on the proton and neutron sharing during the elongation process is put forward. In addition, the measurement of the total kinetic energy at scission allows the determination of the distance between fragments and the total excitation energy available. 

In the low-energy fission of $^{240}$Pu, $\langle TKE^*\rangle$ and the mass difference $\langle Q_\mathrm{gs}\rangle$ show similar structures that may be a signature for the formation of little-deformed, closed-shell nuclei. These structures compensate when subtracted to estimate the total excitation energy $\langle TXE_{\mathrm{gs}}^*\rangle$ released in fission. This leads to an almost constant energy release of about $\langle TXE_{\mathrm{gs}}^*\rangle\sim 25$~MeV over the complete fragment distribution. The symmetric fission is observed to be associated with a slightly more energetic release of almost 30~MeV. 
In the higher-energy fusion-induced fission of $^{250}$Cf, with no neutron evaporation considered between saddle and scission, the neutron excess of the scission fragments shows a steeper change around symmetry when compared to LD-SP and GEF models, resulting in more neutron-rich heavy fragments. The behaviour of $\langle TKE^*\rangle$ points to a smooth evolution of the scission configuration with the fragment split, with a flat behaviour around symmetry. This behaviour, together with the estimation of the distance between the fragments, indicates an almost constant elongation, though different from simple liquid-drop model considerations. Also in $^{250}$Cf higher-energy fission, a strong evolution of $\langle TXE_{\mathrm{gs}}^*\rangle$ is observed, with 25~MeV more excitation energy in the symmetric splits than in the very asymmetric splits. 

A lower limit for the fraction of $TXE^*$ released by neutron evaporation was found to evolve with the fragment split and to be above 50\% of the total in both systems. The calculation of a lower limit of excitation energy released by neutron evaporation as a function of $Z$ suggests that the partition of $TXE^*$ between the fragments according to their masses is not valid for these systems with $E^*_\mathrm{FS}\sim 9$ and $\sim 42$~MeV; being more suitable the description with unbalanced temperatures and continuous flow of energy from the light to the heavy fragment.

\section*{Acknowledgements}
M. Rejmund and A. Navin are acknowledged for their strong support during the experiment. M.C. was financially supported by the Programme ``Ram\'on y Cajal'' of the Spanish Ministry of Economy, under contract RYC-2012-1185. C.R.-T. was financially supported by the Programme ``Axudas de apoio a etapa postdoutoral do plan galego de investigaci\'on, innovaci\'on e crecemento 2011-2015 (Plan I2C)'' of the Xunta de Galicia.

\end{document}